\documentclass{jfm}

\usepackage{amsmath}
\usepackage{color}
\usepackage{amssymb}
\usepackage{graphicx}
\usepackage{upgreek}

\newcommand{\eqr}[1]{(\ref{eq:#1})}

\newcommand{\pd}[2]{\frac{\partial #1}{\partial #2}}

\newcommand{\ie}{\textit{i.e.}}
\usepackage{dcolumn}% Align table columns on decimal point
\usepackage{bm}% bold math
\usepackage{amsmath}
\usepackage{amssymb}
\usepackage{scalerel}[2014/03/10]
\usepackage{stackengine}

\usepackage[euler]{textgreek}

\usepackage{multirow}

\title[ ]{Impact of inlet gas turbulence on the formation, development, and breakup of interfacial waves in a two-phase mixing layer}

\author{D.~Jiang \and Y. Ling\aff{1}
  \corresp{\email{stanley\_ling@baylor.edu}}
}

\affiliation{
\aff{1}Department of Mechanical Engineering, Baylor University, Waco, TX 76798, USA
}

\begin{document}

\maketitle

\begin{abstract}
Understanding the development and breakup of interfacial waves in a two-phase mixing layer between the gas and liquid streams  is paramount to atomization. Due to the velocity difference between the two streams, the shear on the interface triggers a longitudinal instability, which develops to interfacial waves that propagate downstream. As the interfacial waves grow spatially, transverse modulations arise, turning the interfacial waves from quasi-2D to fully 3D. The inlet gas turbulence intensity has a strong impact on the interfacial instability. Therefore, parametric direct numerical simulations are performed in the present study to systematically investigate the effect of the inlet gas turbulence on the formation, development, and breakup of the interfacial waves. The open-source multiphase flow solver, PARIS, is used for the simulations and the mass-momentum consistent volume-of-fluid method is used to capture the sharp gas-liquid interfaces. Two computational domain widths are considered and the wide domain will allow a detailed study of the transverse development of the interfacial waves. The dominant frequency and spatial growth rate of the longitudinal instability are found to increase with the inlet gas turbulence intensity. The dominant transverse wavenumber, determined by the Rayleigh-Taylor instability, scales with the longitudinal frequency, so it also increases with the inlet gas turbulence intensity. The holes formed in the liquid sheet is important to the disintegration of the interfacial waves. The holes formation is influenced by the inlet gas turbulence. As a result, the sheet breakup dynamics and the statistics of the droplets formed also change accordingly. 
\end{abstract}

%Use showkeys class option if keyword
\maketitle

\section{Introduction}
%Define your problem
When a fast gas stream interacts with a parallel co-flowing liquid stream of lower velocity, the gas-liquid interface is unstable and interfacial waves will form and develop. As the interfacial waves are advected downstream, the waves grow in amplitude, deform, and eventually break into small droplets. The droplets are dispersed by the gas stream, forming a two-phase mixing layer between the gas and liquid streams. The  two-phase mixing layer is at the heart of numerous twin-fluid atomization processes, such as air-assisted and air-blast atomizations \citep{Lefebvre_1980a,Lefebvre_1988a}. 

%Streamwise instability
The formation and early development of the interfacial waves are mainly controlled by the longitudinal shear-induced Kelvin-Helmholtz-like instability. The most unstable mode in the longitudinal instability dictates the frequency and the wavelengths of the interfacial waves. Linear stability analysis has been carried out to predict the most-unstable longitudinal wave frequency $f$ and the wavelength $\lambda$. While inviscid analysis  \citep{Raynal_1997a, Marmottant_2004a, Matas_2011a} yielded reasonable predictions for the frequency but under-predicted the spatial growth rate, viscous temporal analysis \citep{Boeck_2005a} well predicted the growth rate but overestimated frequency. It was later shown by \cite{Otto_2013a} and \citet{Fuster_2013a} that viscous spatial-temporal stability analysis is required to well predict both the frequency and the growth rate. Due to the velocity deficit induced by the wake of the separator plate, the instability can transfer from convective to absolute instability regimes. When the instability is absolute, a dominant unstable mode will arise. \cite{Matas_2015b} further showed that the confinement effect induced by the stream thickness needs to be taken into account to yield a good prediction of the most unstable mode. {When the longitudinal wavelength $\lambda$ is significantly lower than the stream thickness, then it scales with the gas stream boundary layer thickness $\delta$ at the inlet and the absolute instability is mainly controlled by the surface tension mechanism. In contrast, if the most-unstable wavelength is comparable or higher than the stream thickness, then the inviscid confinement absolute instability will become dominant \citep{Matas_2018a}. In such a case, the wave propagation speed was found to agree well with the Dimotakis speed \citep{Dimotakis_1986a}, $U_D=(\sqrt{\rho_l}U_l+\sqrt{\rho_g}U_g)/(\sqrt{\rho_l}+\sqrt{\rho_g})$, where $\rho_l,U_l$ and $\rho_g,U_g$ represent the densities and velocities for the gas and liquid streams, respectively.} The most unstable wavelength and frequency can thus be related to each other by the Dimotakis speed as {$\lambda= U_D/f$}. 

%Impact of inlet gas turbulence on streamwise instability
Conventionally, the stability analysis and numerical studies of the longitudinal shear-induced instability assume both the gas and liquid streams are laminar when they meet \citep{Boeck_2005a, Otto_2013a, Fuster_2013a, Agbaglah_2017a, Ling_2017a, Ling_2019a}. Nevertheless, turbulent fluctuations may exist in the gas stream in experiment due to the high Reynolds numbers. As indicated by \cite{Matas_2015a} that, this may be a potential reason for the discrepancies between different experiments. The impact of the gas inlet turbulence on the longitudinal instability has been recently investigated  through experiment by \cite{Matas_2015a} and later using direct numerical simulation by \cite{Jiang_2020a}. Both the frequency and the spatial growth rate were observed to increase with the inlet gas turbulence intensity $I$, when $I$ is over a threshold \citep{Matas_2015a, Jiang_2020a}. Attempts have been made to incorporate the effect of inlet gas turbulence in the linear viscous spatial-temporal stability analysis based on turbulent viscosity models. The modified stability analysis reasonably capture the trend, \ie, the frequency increases with $I$, but underestimates the values \citep{Jiang_2020a}. An accurate linear stability theory that captures the most unstable modes for a turbulent gas stream remains to be developed.

%Transverse instability and interfacial wave development; impact of inlet gas turbulence. 
Transverse modulations on the longitudinal waves develop when the waves grow and propagate downstream. The Rayleigh-Taylor (RT) instability has been shown to be the primary driving mechanism for the transverse  instability in a cylindrical coaxial configuration \citep{Varga_2003a, Marmottant_2004a}. The azimuthal wavelength was estimated based on inviscid RT instability on a planar surface with undulations from the longitudinal instability. When the interface accelerates toward liquid or decelerates toward gas, the interface is unstable and the most-unstable wavelength is dictated by the surface tension and the interface acceleration. In the work of \citet{Marmottant_2004a}, the maximum acceleration is estimated based on experimental correlation and is a function of $\text{We}_\delta$, a Weber number  based on $\delta$. 
Numerical studies by \citet{Jarrahbashi_2014a} showed that both the RT instability (baroclinic effects) and strain-vorticity interaction contribute to the transverse instability development. The latter is more important when the gas-to-liquid density ratio is high. The sequential development of the transverse variation of the interfacial wave is influenced by the interaction between the interfacial waves and the gas stream. As the interfacial waves grow and invade into the gas stream, the accelerated gas flow above the wave crest induces a Bernoulli depression, which enhances the growth of the wave \citep{Hoepffner_2011a}. Since the transverse instability is closely tied to the longitudinal counterpart and the latter is in turn influenced by the inlet gas turbulence, it is expected that the inlet gas turbulence will also play a significant role in the transverse instability and the subsequent transverse development of the interfacial waves. A detailed analysis on the effect of inlet gas turbulence on transverse instability features such as  the dominant transverse wavenumber remains absent in the literature. 

There exist multiple pathways for the 3D interfacial waves to break into filaments and droplets: the fingering and the hole-in-sheet modes. Visualization of these two breakup modes have been presented in high-fidelity numerical simulations \citep{Jarrahbashi_2016a, Ling_2017a, Zandian_2018a}. The fingering modes typically occur when the disintegration of the interfacial wave is relatively mild. In such a case, the Rayleigh-Plateau (RP) instability gets a chance to develop at the Taylor-Culick rims on the edge of the liquid sheets extended from the waves \citep{Roisman_2010a,Agbaglah_2013a}. The RP instability results in liquid fingers which are approximately aligned with the streamwise direction. The number of fingers formed is related to the dominant transverse wavenumber \citep{Marmottant_2004a}.  The liquid fingers will continue to break into droplets.

The formation of droplets due to pinching of a filament is by itself a complicated subject \citep{Eggers_1993a, Ambravaneswaran_2002a,Castrejon-Pita_2015a,Zhang_2019b}. In general, the primary droplets formed are similar to the local diameter of the filament, while the secondary satellite droplets can be much smaller. Since the liquid fingers typically exhibit irregular shapes, the breakup dynamics is more complex than the classic Rayleigh breakup of a liquid cylinder and generally leads to a distribution of droplets of different sizes \citep{Villermaux_2004a, Ling_2017a}. The size distribution of droplets formed is essential to spray applications and different distribution models have been proposed, \eg, the log-normal,  exponential, Poisson, Weibull-Rosin-Rammler, Pareto, and gamma distributions \citep{Villermaux_2004a, Herrmann_2011a, Marty_2015a, Ling_2017a,Kooij_2018a, Balachandar_2020a}. These distribution functions agree with different sets of experimental or simulation data in some extent. Whether there exists a universal distribution of droplet size remains an unresolved question. 

When long liquid sheets extend from the interfacial wave crest and has a strong interaction with the gas stream, the disintegration of the interfacial wave is generally more violent and tends to follow the hole-in-sheet mode \citep{Ling_2017a}. The hole expansion speed follows the Taylor-Culick velocity \citep{Opfer_2014a, Marston_2016a, Ling_2017a, Agbaglah_2021a}. The holes grow and merge, and eventually lead to a violent rupture of the sheet, producing separate ligament, droplets of different sizes and orientations, and fingers that remain attached to the liquid sheet. The formation of a hole in a liquid sheet is due to the pinching of the two surfaces of the liquid sheet, similar to the pinching of a filament in drop formation. The pinching of surfaces is induced by the disjoining pressure when the distance between the two surfaces is sufficiently small (O(10 nm)). In numerical simulations, the cell size is usually much larger, so the disjoining pressure is typically not included in the physical model. As a result, the minimum cell size serves as the numerical cut-off length scale to pinch a liquid sheet and to form holes. The absence of disjoining pressure seems to have little effect on the surface pinching, since the process for low-viscosity liquids is mainly dictated by the fluid inertia, similar to the droplet formation due to the pinching of filament \citep{Zhang_2019b}. Nevertheless, a careful grid-refinement study is still required to verify the simulation results, in particular for the statistics of the droplets formed \citep{Ling_2017a}. Former studies on the interfacial waves breakup assume that the inlet gas stream is laminar. The effect of the inlet gas turbulence on the sheet breakup dynamics and the droplets statistics remains unclear. 

%Goal of study
The goal of the present study is to investigate the fate of the interfacial waves in a two-phase mixing layer between parallel liquid and gas streams through direct numerical simulations. As an extension of our former studies \citep{Ling_2017a, Ling_2019a, Jiang_2020a}, the present study is focused on the transverse interfacial instability and the impact of inlet gas turbulence on the formation, development, and breakup of the three-dimensional interfacial waves. To allow a detailed investigation of the transverse features of interfacial waves, we have used a computational domain that is three times as wide as that in our former studies. The key questions we aim to address include: 
\begin{enumerate}
\item What are the physical mechanisms that drive the transverse development of the interfacial waves? Can one predict the transverse wavenumber based on the stability theory? 
\item How will the inlet gas turbulence influence the longitudinal and transverse interfacial instability and the development of the 3D interfacial waves? 
\item What are the pathways for the interfacial waves to disintegrate into filaments and droplets? What is the effect of the inlet gas turbulence on the interfacial wave breakup dynamics and the statistics of the droplets formed? 
\end{enumerate}

The rest of the paper will be organized as follows. The problem description and the simulation approaches, including the governing equations, the numerical methods, and the simulation setup, will be presented in section \ref{sec:methods}. The simulation results will be shown in section \ref{sec:results}. The longitudinal and transverse instabilities of the interfacial waves, the development and breakup of the interfacial waves, and the droplet statistics will be discussed in sequence. The key conclusions of the present study will be summarized in section \ref{sec:conclusions}.

\section{Simulation Methods}
\label{sec:methods}

\subsection{Problem description}
\begin{figure}
\centering
\includegraphics[trim={0cm 0cm 0cm 0cm},clip,width=.95\textwidth]{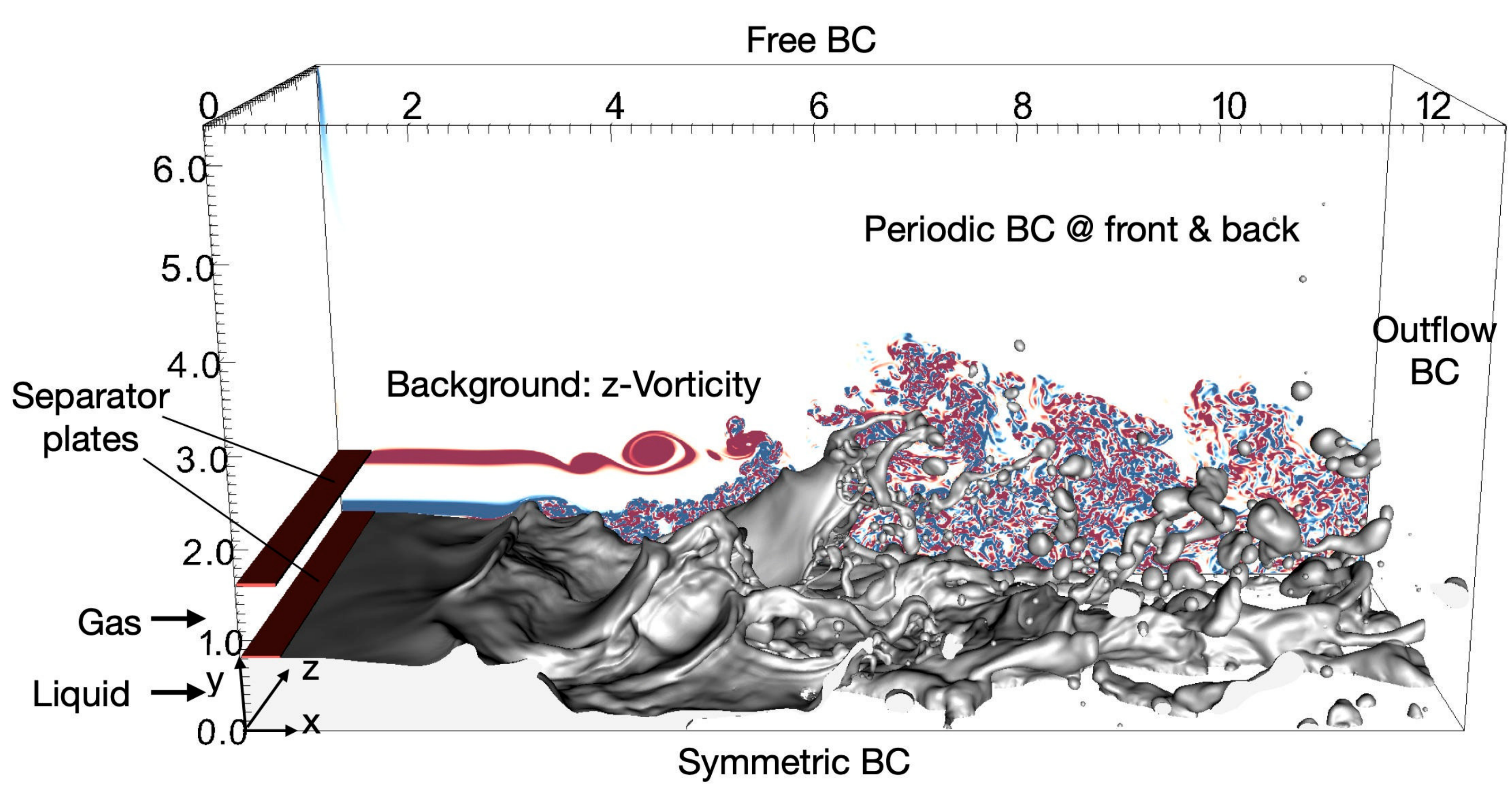}
\caption{Simulation setup for the interfacial waves between parallel liquid and gas streams. }
\label{fig:sim}
\end{figure}

The two-phase mixing layer to be considered in the present study is illustrated in figure\ \ref{fig:sim}. Two parallel planar gas and liquid streams enter the domain from the left. Two thin solid plates are placed near the inlet to separate the two streams, mimicking the injector nozzle. The inclusion of the separator plate has been shown to be important to the interfacial instability \citep{Otto_2013a}. The two streams meet at the end of the lower separator plate. The liquid inflow is laminar, while turbulent velocity fluctuations of different intensity levels are present at the gas inlet. The mean gas velocity is significantly higher than that for the liquid. As a result, the gas-liquid interface is unstable, and wavy structures develop on the interface and propagate downstream. For convenience of discussions, we refer to the $x$, $y$, and $z$ directions as the {longitudinal, vertical, and transverse directions}, respectively. The interfacial waves, when they are just formed, are approximately 2D and longitudinal. Yet as they grow and propagate downstream, transverse modulations arise and the waves evolve to be fully 3D. The interfacial waves will interact with the gas stream as the amplitudes become finite. The eventual breakups of the interfacial waves produce small ligaments and droplets that are mixed with the gas stream, forming a two-phase mixing layer. After an initial transition period for the two streams to progressively enter the domain, the wave formation and the two-phase turbulent flow reaches a statistically stationary state \citep{Ling_2017a, Ling_2019a}. 

\subsection{Governing equations}
The two-phase interfacial flows are governed by the incompressible Naviers-Stokes equation with surface tension. The one-fluid approach is employed, where the gas and liquid phases are treated as one fluid with material properties change abruptly across the interface. The momentum and continuity equations are expressed as 
\begin{align}
	\rho \left(\pd{ u_i}{t} + u_i \pd{u_j}{x_j}\right) & = -\pd{p}{x_i} + \pd{}{x_j}\left[ \mu\left( \pd{u_i}{x_j} + \pd{u_j}{x_i} \right) \right] + \sigma \kappa\delta_s n_i\, ,	
	\label{eq:mom}\\
	\pd{u_i}{x_i} & =0\, .	
	\label{eq:cont}
\end{align}
where $\rho, u_i, p, \mu$ represent density, velocity, pressure and viscosity, respectively.  The last term on the right hand side of the momentum equation represents the surface-tension, which is a singular force localized on the sharp interface using the Dirac distribution function $\delta_s$. The surface tension coefficient $\sigma$ is taken to be constant, and $\kappa$ and $n_i$ represent the curvature and normal vector of the interface. 

The gas and liquid phases are distinguished by a characteristic function $\chi$, while $\chi=1$ and $0$ represent the liquid and gas phases, respectively. The interface evolution can be captured by solving the advection equation of $\chi$. 
\begin{align}
  \pd{\chi}{t} + u_i \pd{\chi }{x_i} & = 0 \, .
  \label{eq:adv}
\end{align}

The mean value of $\chi$ in a computational cell is defined as 
\begin{align}
	f = \frac{1}{\Delta\Omega}\int_{\Omega} \chi dV\, .
	\label{eq:liq_vof}
\end{align}
which also represents the volume fraction of liquid ($\chi=1$) in a cell. Correspondingly, the gas volume fraction in a cell is $\hat{f}=1-f$. The fluid properties in interfacial cells with $0<f<1$ are calculated based on arithmetic mean 
\begin{align}
	\rho = f \rho_l + \hat{f} \rho_g\,, \quad
	\mu =f\mu_l + \hat{f}\mu_g\,.
	\label{eq:mix_density}
\end{align}
where the subscripts $l$ and $g$ represent variables corresponding to the liquid and gas phases, respectively. 

\subsection{Numerical methods}
The governing equations are solved by the finite volume method on a staggered grid. The advection equation, Eq.~\eqr{adv}, is solved using a geometric volume-of-fluid (VOF) method. The interface normal is computed  following the mixed Young's-centred method of \citet{Aulisa_2007a}. The Lagrangian-explicit scheme of \citet{Li_1995b} is used for the VOF advection \citep{Scardovelli_2003a}.  The convection term in the momentum equation, Eq.\eqr{mom}, is discretized consistently with the VOF method \citep{Arrufat_2020a}, and this mass-momentum consistence has been shown to be crucial in capturing interfacial dynamics when large velocity and density contrasts are present at the interface \citep{Rudman_1998a, Ling_2017a,Vaudor_2017a, Zhang_2020a, Arrufat_2020a}. The incompressibility condition is incorporated using the projection method \citep{Chorin_1968a}. The pressure Poisson equation is solved using PFMG multigrid solver in the HYPRE library. The viscous term is discretized explicitly using the second-order centered difference scheme. The interface curvature is calculated using the height-function method of \citet{Popinet_2009a} and the balanced continuous-surface-force method is used to discretize the surface tension term  \citep{Renardy_2002a, Francois_2006a, Popinet_2009a}. The time integration is done by a second-order predictor-corrector method.  To capture the dynamics of under-resolved droplets less erroneously than by just quasi-fragment VOF patches, droplets of size smaller than about two cells are converted into Lagrangian point-particles and are traced under the one-way coupling approximation, following the approach of \citet{Ling_2015a}.

The aforementioned numerical methods have been implemented in the open-source solver, \emph{PARIS-Simulator}. Detailed implementations and validation of the code can be found in previous studies \citep{Tryggvason_2011a, Ling_2015a, Ling_2017a, Ling_2019a, Arrufat_2020a, Aniszewski_2021a}.

\subsection{Simulation setup}
%(XXX BC for the domain, separator plate, cell size, time step, justification of setup. Physical time range of simulation, why? Computation cost, cores, computers, cpu-time.  A table for all simulation cases and define case names. XXX) 
The computational domain is a cuboid. The dimensions in $x$ and $y$ directions for the two domains are the same, \ie, $L_x=16H$ and $L_y=8H$, where $H$ is the height of the liquid stream at the inlet.  Periodic boundary conditions are applied to the front and back of the domain. A free boundary condition is applied at the top, so gas is allowed to freely flow through the boundary. The velocity outflow boundary condition is imposed at the right surface. {The bottom surface is treated as a slip wall to avoid the modeling challenge for moving contact lines on a no-slip surface. The contact angle is specified as 90 degrees on the bottom surface. As a result, the bottom is equivalent to a symmetric boundary and the present simulation setup can be viewed as a symmetric model for the airblast atomization configurations with gas streams on both sides of the liquid stream \citep{Chaussonnet_2020a}, which could capture the interfacial stability when the symmetric mode is dominant. A full simulation with both gas streams will be required to resolve the asymmetric instability mode and breakup dynamics \citep{Delon_2018a}.} 

The length and height of the domain and boundary conditions used have been examined in previous studies, which are shown to be sufficient to resolve present problem \citep{Ling_2019a}. Two different domain widths, $L_z=2H$ and  $6H$,  are considered. Some of the simulation results for the narrow domain have been shown in our previous study \citep{Jiang_2020a}. While the narrow domain is useful in capturing the longitudinal wave formation, the wide domain ($L_z=6H$) is required to investigate the transverse instability and the transverse development of the interfacial waves.  

The gas stream has a similar thickness as the liquid stream, \ie, $H-\eta_y$, where $\eta_y$ is the thickness of the thin separator plate. The liquid and gas properties are similar to those of water and pressurized air, following the previous studies \citep{Ling_2017a, Ling_2019a, Jiang_2020a}, see table \ref{tab:phy_para1}. 

The mean flow at the inlet is horizontal, so the $y$- and $z$-components of the mean velocity are zero, \ie, $\overline{v}_0=\overline{w}_0=0$. The $x$-component of the mean velocity at the inlet is expressed as
\begin{equation}
	\overline{u}_{0}(y) = 
	\left \{
		\begin{array}{ll}
		U_{l}\,\mathrm{erf} \left[\frac{H-y}{\delta}\right], &  0 \leq y < H\ \text{(liquid stream)}, \\
		0,                                               &  H \leq y < H + \eta_y\ \text{(separator plate)}, \\
		U_{g}\,\mathrm{erf}\left[\frac{y-(H+\eta_y)}{\delta}\right]\mathrm{erf}\left[\frac{2H-y}{\delta}\right], & H+\eta_y \leq y < 2H\ \text{(gas stream)}, \\
		0, &  \mathrm{else}\ \text{(wall)}. 
		\end{array} 
	\right.
\end{equation}
The velocities in the gas and liquid stream are generally uniform and are equal to $U_l$ and $U_g$ away from the separator plates, respectively, see figure \ref{fig:inlet_profile}(a). The error function is used to model velocity profile in the boundary layers near the separator plates. The parameter $\delta$ characterizes the boundary layer thickness, which is taken to be the same for both the gas and liquid streams, \ie, $\delta=H/8$. The dimensions of the two separator plates are the same, the thickness and length of which are $\eta_y=H/32$ and  $\eta_x=H/2$, respectively. The thickness $\eta_y$ is chosen to be significantly smaller than $\delta$. Based on the former study of \cite{Fuster_2013a}, the specific value of $\eta_y$ has negligible effect on the interfacial instability. 
The values for the key physical parameters for the present problem are listed in table \ref{tab:phy_para1}.

\begin{table}
  \begin{center}
  \begin{tabular}{ccccccccc} 
   $\rho_{l}$      & $\rho_{g}$     & $\mu_{l}$ & $\mu_{g}$ & $\sigma$  & $U_{l}$ & $U_{g}$ & $H$ & $\delta$\\ 
   (kg/m$^{3}$) & (kg/m$^{3}$) & (Pa\,s) & (Pa\,s)  & (N/m) & (m/s) & (m/s) & (mm)	& (mm)\\
   1000             & 50                  & $\mathrm{10^{-3}}$ & 5$\times10^{-5}$ & 0.05 & 0.5 & 10 & 0.8 & 0.1\\ 
  \end{tabular}
  \caption{Physical parameters.}
  \label{tab:phy_para1}
    \end{center}
\end{table}

Pseudo turbulent velocity fluctuations are superposed on the gas velocity at the inlet (\ie, $H+\eta_y \leq y < 2H$) as 
\begin{equation}
	u_{0} = \overline{u}_{0} + u'(y,z,t)\, ,\quad 	v_{0} = v'(y,z,t)\, ,\quad 	w_{0} = w'(y,z,t)\, ,
\end{equation}
where the turbulent fluctuations $u', v', w'$ are computed using the digital filter approach of \citet{Klein_2003a}. The filter length is $1/4H$. Implementation details and verification for the approach to generate turbulent velocity fluctuations have been given in previous study \citep{Jiang_2020a} and thus are not repeated here. 

\subsection{Data collection and processing}
%(XXX Define time averaging $\overline{a}$, time and space averaging $\langle a \rangle$. Explain interfacial height measurement based on height function. XXX) 
Time and spatial averaging are performed to process the instantaneous 3D field of simulation data. The time averaging operator for a variable $a$ is denoted  as
\begin{align}
	\overline{a}= \frac{1}{t_1-t_0}\int_{t_0}^{t_1}a(t) \, dt\,
	\label{eq:liq_vof}
\end{align}
where $t_0$ and $t_1$ represent the beginning and the end time of the sampling. Sampling of data will not start until the simulation reaches the statistically steady state at about $t^*=200$. 

The spatial averaging in the transverse $z$-direction is defined as 
\begin{align}
	\langle a\rangle =  \frac{1}{L_z} \int_{0}^{L_z}a(z) \, dz 
	\label{eq:liq_vof}
\end{align}
The operation of double temporal and spatial averaging is denoted as $\langle \overline{a}\rangle$.

%For some variables, we have performed averaging both in time and in the transverse $z$ direction. The double  averaging operator is defined as
%\begin{align}
%	\langle a \rangle = \frac{1}{t_1-t_0}  \frac{1}{L_z} \int_{t_0}^{t_1}\int_{0}^{L_z}a(z,t) \, dz \, dt
%	\label{eq:liq_vof}
%\end{align}

The height functions, which are computed for the interface curvature, will also be used to evaluate the interface location. The interfacial height, \ie, the $y$-coordinate of the interfacial location, is denoted by $h(t,x,z)$. The root-mean-square of the interfacial height fluctuations, $\overline{h'h'}$, represents the thickness of the two-mixing layer or the amplitude interfacial wave, where $h'=h-\overline{h}$.

\subsection{Key dimensionless parameters}
%(XXX Define characteristic velocity and length scales. Define the key dimensionless parameters and explain their effect on the problem. Where is the current case located in the parametric regime. {Define dimensionless variables $t^*=tU_g/H$, $u^*=u/U_g$, $x^*=x/H$ and so on and use only dimensionless variables in the rest of the paper consistently.}   XXX)
With $H$ and $U_g$ as the scaling variables, the dimensionless time, velocity, and length are defined as $t^*=tU_g/H$, $u^*=u/U_g$, $x^*=x/H$, respectively. The key parameters listed in table \ref{tab:phy_para1} can be converted to the dimensionless form, see table \ref{tab:dmls_para}. In addition to $H$, the boundary layer thickness $\delta$ can also be important to the interfacial instability  \citep{Otto_2013a}.  As a result, two different Reynolds numbers, \ie, $Re_{g,H}$ and $Re_{g,\delta}$, are defined correspondingly. Due to the high $\mathrm{Re}_{g,H}$, the gas stream will be turbulent even if the gas inflow is laminar. The Weber number based on $\delta$ characterize the effect of surface tension on the interfacial instability development \citep{Otto_2013a}. The gas-to-liquid dynamic pressure ($M$) is important in determining the macro-scale features, such as the breakup length \citep{Lasheras_1998a}, and also the {regime for the shear-induced longitudinal instability} (\cite{Otto_2013a, Fuster_2013a}). 
{According to the previous studies \citep{Ling_2019a,Jiang_2020a}, the selected dimensionless parameters place the shear longitudinal instability in the absolute regime. There are two different mechanisms, namely surface-tension and confinement, that can drive a transition from convective to absolute regimes. Based on the results to be shown later, the absolute longitudinal instability in the present problem seems to belong to the confinement category.}

\subsection{Simulation cases}
\begin{figure}
\centering
\includegraphics[trim={0cm 0 0cm 0},clip,width=.95\textwidth]{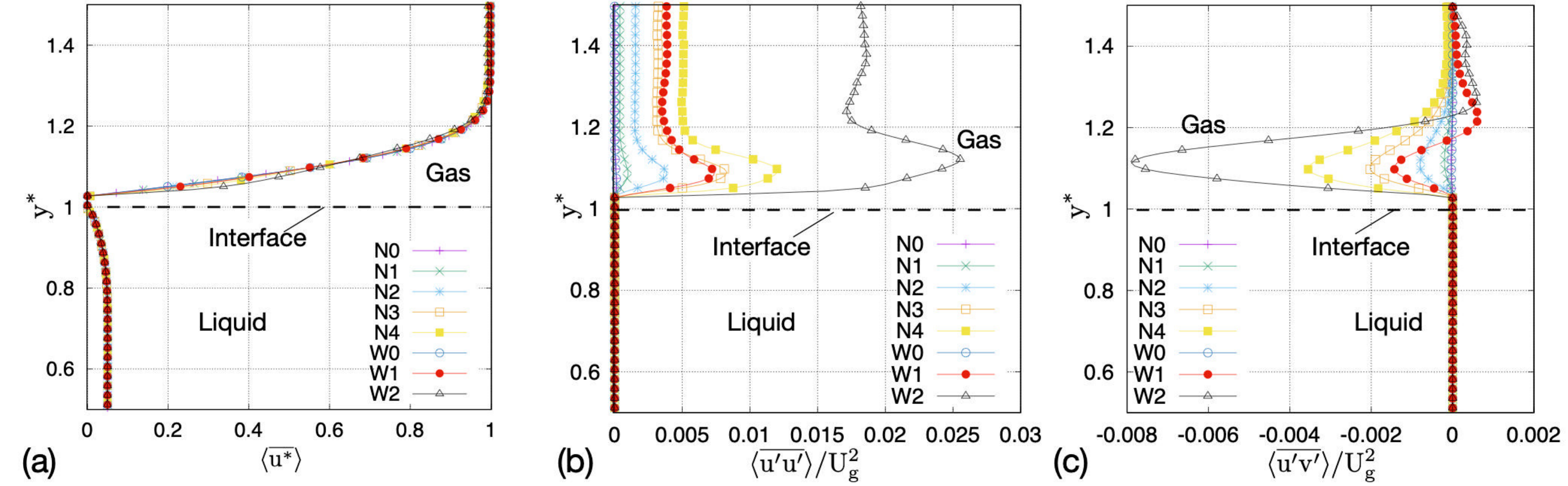}
\caption{Temporally and spatially (in $z$-direction) averaged profiles of (a): velocity in streamwise direction (b): normal Reynolds stress and (c): shear Reynolds stress.}
\label{fig:inlet_profile}
\end{figure}

A parametric study is carried out to systematically investigate the inlet gas turbulence intensity $I$. The simulation cases are summarized in table \ref{tab:simu_runs}. For the narrow domain, five different $I$ are considered (cases N0 to N4), while for the wide domain, only three different $I$ (cases W0 to W2) are simulated due to the higher computational costs.

\begin{table}
  \centering
  \begin{tabular}{ccccccc} 
    $M$ & $r$         
& $m$  & $\mathrm{Re}_{g,\delta}$ & $\mathrm{We}_{g,\delta}$ & $\mathrm{Re}_{g,H}$ & $I$\\[2pt]
  $\rho_g U_g^2/(\rho_l U_l^2)$   &  $\rho_l/\rho_g$ 
& $\mu_l/\mu_g$  &  $\rho_g U_g \delta/\mu_g$  &   $\rho_g U_g^2 \delta/\sigma$ & $\rho_g U_g H_g/\mu_g$ & $(\sqrt{\overline{u'u'}})_e/U_g$ \\[4pt]
  20 & 20 & 20 & 1000 & 10 & $7750$ & 0, 0.06, 0.13 \\
  \end{tabular}
  \caption{Key dimensionless parameters.}
  \label{tab:dmls_para}
\end{table}

Temporally and spatially averaged profiles of the streamwise velocity ($\langle \overline{u}\rangle/U_g$), normal and shear Reynolds stresses ($\langle \overline{u'u'}\rangle/U_g^2$ and $\langle \overline{u'v'}\rangle/U_g^2$) at the end of the separator plate ($x=\eta_x$) for all the cases are shown  in figure \ref{fig:inlet_profile}. It can be seen that the mean velocity profiles for all the cases are almost identical. For the cases N0 and W0, the Reynolds stresses are negligibly small. Magnitudes of the Reynolds stresses generally increase with {the inlet gas turbulence intensity}. The profiles of the normal and shear stresses are consistent with those for planar turbulent channel flows, indicating the turbulent fluctuations generator at the inlet has been implemented properly \citep{Klein_2003a}. The normal Reynolds stress ($\langle \overline{u'u'}\rangle/u_g^2$) increases with $y^*$ from zero at the separator plate and reaches the maximum at around the middle of the boundary layer. Then it decreases and approaches a constant outside of the boundary layer, the square root of the normal Reynolds stress,\ie, $I=(\langle \overline{u'u'}\rangle_{y^*=1.5})^{1/2}/U_g$, is used to characterize the inlet gas turbulence intensity.

The computational domains are discretized using a fixed regular cubic grid. The cell size  $\Delta=6.25$ \textmu m ($H/\Delta =128$) is used for all the cases. The cell size has been verified to be adequate for good estimates of high-order two-phase turbulence statistics, such as TKE dissipation \citep{Ling_2019a}. The numbers of cells for the narrow and wide domains are about 0.5 and 1.6 billions, see table \ref{tab:simu_runs}.  All cases are run for a physical time to at least $t^*=450$, which is sufficient to cover the formation of about 12 to 18 waves after the statistically steady state is reached. The present simulations are performed on the Intel Xeon Platinum 8160 (Skylake) computing nodes on the TACC Stampede2 machine. For the narrow domain cases (N0 to N4), 22 nodes are used, while 64 nodes are used for the wide domain cases (W0 to W2). The total computing time used for all the cases is about 280,000 node-hours (13,400,000 core-hours).

\section{Results and Discussion}
\label{sec:results}
\subsection{General behavior}
\begin{figure}
\centering
\includegraphics[trim={0cm 0cm 0cm 0cm},clip,width=1\textwidth]{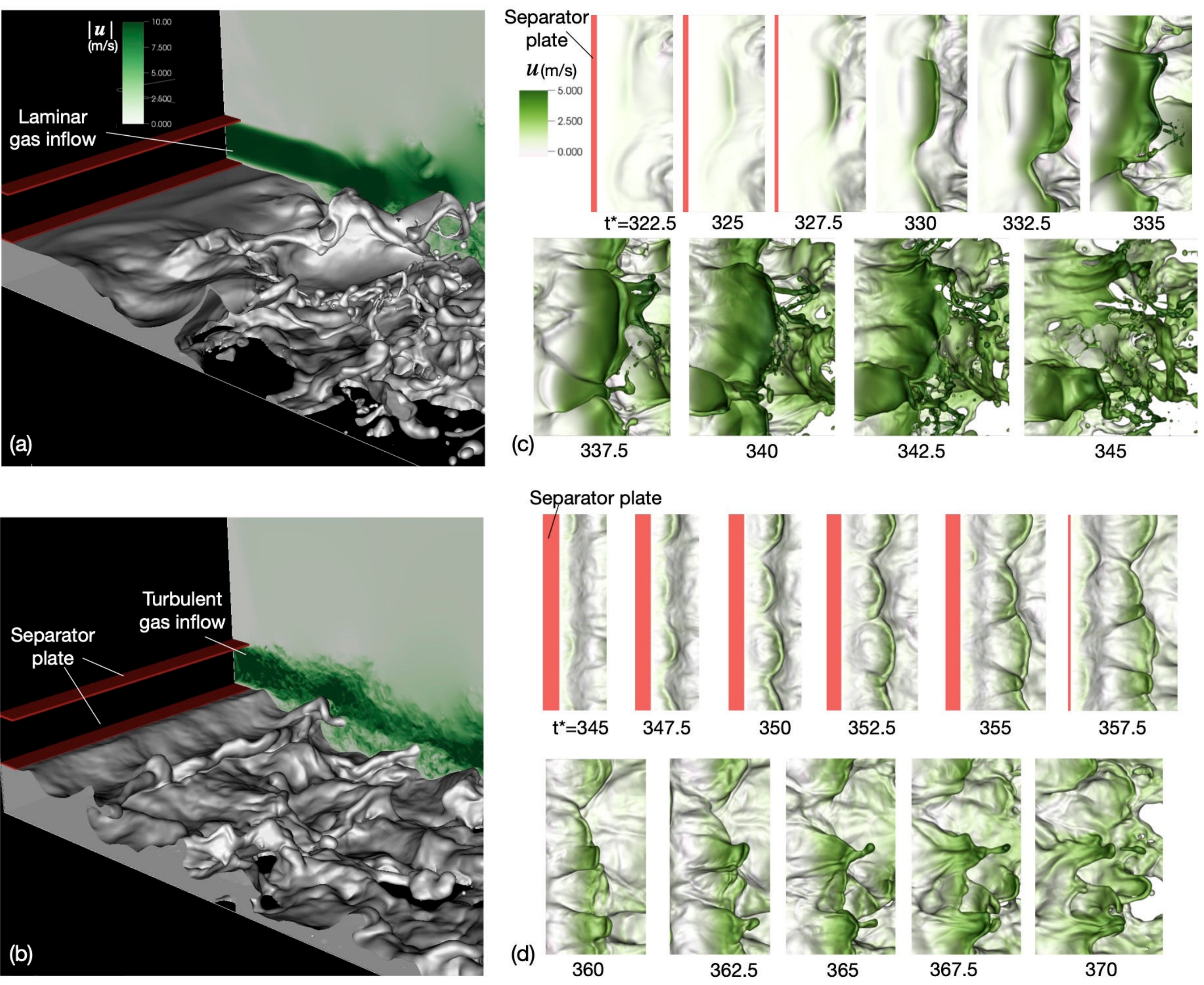}
\caption{{Snapshots for the two-phase mixing layer for (a) the case W0 with a laminar gas inflow and (b) the case W2 with a turbulent gas inlet. The color on the background in (a) and (b) represents the velocity magnitude. Temporal development of the interfacial waves for the cases W0 and W2 are shown in (c) and (d), respectively, where the color on the interfaces represents the streamwise velocity. The frames in (c) and (d) are moving with the waves to keep the waves located at the center of the window.} Supplementary movies are available online. }
\label{fig:general_behavior}
\end{figure}

The temporal evolutions of the interfacial waves and the velocity fields for the cases W0 and W2 are shown in figure \ref{fig:general_behavior}. The cases W0 {and} W2 represent the two distinct gas inflow conditions: laminar ($I=0$) and highly turbulent ($I=0.13$). The figures {\ref{fig:general_behavior}(a) and \ref{fig:general_behavior}(b)}  in the left column show the interfacial waves from a 3D view, with the streamwise $u$-velocity on the background. The figures {\ref{fig:general_behavior}(c) and \ref{fig:general_behavior}(d)} in the right column are sequential snapshots of the gas-liquid interfaces colored by the $u$-velocity from the top view. 

The results for the case W0 shown in figure \ref{fig:general_behavior}{(c)} represent a typical process of interfacial wave formation and development for a planar two-phase mixing layer \citep{Matas_2011a, Agbaglah_2017a, Ling_2017a, Zandian_2018a, Ling_2019a}. When the two streams meet at the end of the separator plate, the longitudinal instability is triggered due to the shear at the interface. The instability develops to a 2D longitudinal interfacial wave which propagates downstream, the wave crest of which is approximately a straight line from the top view. Due to the transverse instability,  the height of the wave crest varies over the transverse direction. When the wave amplitude grows, the wave interacts with the fast gas stream and develop into a 3D wave, exhibiting a lobe shape. The liquid lobe later extends to form a liquid sheet which bends and folds. The thickness of the sheet reduces rapidly and unevenly due to the complex sheet deformation. Multiple holes are formed near the edge of the sheet. The expansion and merging of the holes disintegrate the liquid sheet violently. Multiple breakup events occur (see snapshots between $t^*=335$ and 345 {in figure \ref{fig:general_behavior}(c)}), until the sheet completely breaks into droplets. 

\begin{table}
  \centering
  \begin{tabular}{cccccc} 
   Case & $I$  & $\Delta$ (\textmu m)  & $H/\Delta$ & $cells \#$ & $Cores \#$ \\
    W0 & $0.0$    & \multirow{8}{*}{$6.25$ } & \multirow{8}{*}{$128$} & $1.6 \times 10^9$ & $3072$ \\
    W1 & $0.06$  & & & $1.6 \times 10^9$ & $3072$ \\
    W2 & $0.13$  & & & $1.6 \times 10^9$ & $3072$ \\
    N0 & $0.0$     & & & $0.5 \times 10^9$ & $1024$ \\
    N1 & $0.02$   & & & $0.5 \times 10^9$ & $1024$ \\
    N2 & $0.039$ & & & $0.5 \times 10^9$ & $1024$ \\
    N3 & $0.056$ & & & $0.5 \times 10^9$ & $1024$ \\
    N4 & $0.071$ & & & $0.5 \times 10^9$ & $1024$ \\
    \end{tabular}
  \caption{Summary of simulation runs. {Case names starting} with $W$ and $N$ represent the wide and narrow domains, respectively. }
  \label{tab:simu_runs}
\end{table}

By adding small-amplitude turbulent velocity fluctuations at the gas inlet (case W2), several important differences are observed. First of all, the longitudinal instability grows faster. As a result, the wave amplitude near the inlet is much higher than that for the laminar gas inflow. Second, the liquid sheet extended from the wave are much shorter, and experiences a weaker interaction with the gas stream. Third, the wavenumber of the transverse modulation of the interfacial wave increases. While there are about one to two transverse waves shown in figure \ref{fig:general_behavior}{(c)} (see $t^*=330$), about three to four waves are observed in figure \ref{fig:general_behavior}{(d)} (see $t^*=350$). Due to the increase of wavenumber, the width of the lobes formed is reduced and the shape of the rim formed at the edge of the liquid sheet is less regular. Fourth, the breakup dynamics of the interfacial waves also changes significantly. Fewer holes are seen for the case W2. The disintegration of the wave takes a different path, \ie, forming fingers at the rim. Finally, the different breakup dynamic impacts the statistics of the droplets generated. Significantly fewer droplets are formed. Detailed quantitative analysis for these effects will be presented in the following sections.

\subsection{Longitudinal instability and wave formation}
\label{sec:long_instability}

\begin{figure}
\centering
\includegraphics[trim={0cm 0cm 0cm 0cm},clip,width=.95\textwidth]{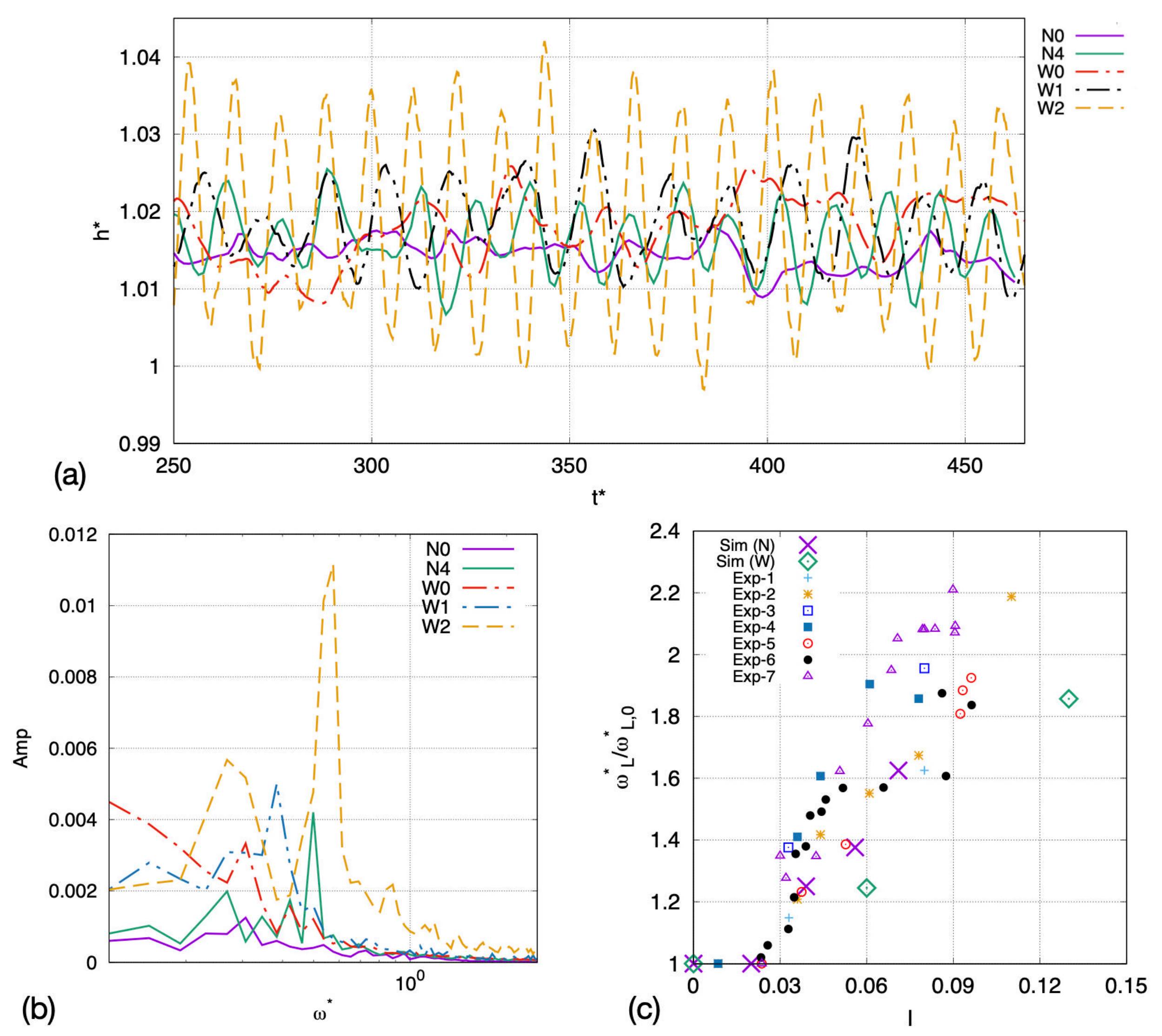}
\caption{(a) Temporal evolution of the {transversely averaged interfacial heights} at $x^{*}=0.625$ for different cases. (b) {Frequency spectra} that correspond to {temporal signal} in (a). (c) {Variation of the normalized dominant frequency for the longitudinal instability, $\omega_L^*/\omega_{L,0}^*$ as a function of $I$ for different cases}, {compared with the experimental data of \cite{Matas_2015a}}.}
\label{fig:temporal_instability}
\end{figure}

{\subsubsection{Absolute instability}}
The shear between the gas and liquid streams triggers a Kelvin-Helmholtz (KH) longitudinal instability, which is the onset of the formation of interfacial waves. The temporal evolutions of the interfacial {heights are} measured near the inlet at $x^{*}=0.625$ and eleven evenly spaced locations along the transverse direction. The temporal evolutions of the spatially-averaged {$h^*=h/H$} are shown in figure \ref{fig:temporal_instability}(a) for different cases of the wide and narrow domains.  At this $x$ location, the wave amplitude {is} small (less than 4\% of $H$ for all cases), so the waves remain in the linear regime. Fourier transform is used to cast the temporal signals to the frequency spectra, see figure \ref{fig:temporal_instability}(b). The spectra obtained at eleven different transverse locations are averaged. The amplitude at the dominant frequency is the maximum in the spectra for all cases except W0, for which the dominant frequency is a local maximum. The maximum amplitude {for the case W0 is located at a very low frequency that} corresponds to the total simulation time. 

{
The appearance of the dominant mode affirms that the cases studied are in the absolute instability regime. The frequency of the longitudinal wave formation is dictated by the most-unstable mode of the longitudinal instability. As suggested by \citet{Matas_2015b} and \citet{Matas_2018a}, there are two different mechanisms that contribute to the convective-to-absolute instability transition: the surface tension and the confinement. Both types of absolute instabilities have been identified in the spatial-temporal stability analysis, as the pinching between the shear branch and the other branch controlled by surface tension \citep{Otto_2013a} or by confinement \citep{Juniper_2006a, Healey_2007a}. For both types of absolute instabilities, the imaginary part of the frequency is positive at the pinch point. The wavenumber at the pinch point for the confinement type is generally low, so that the most-unstable wavelength is typically larger than the layer thickness $H$. In contrast, for the surface-tension type, the wavenumber for the dominant mode is much higher and the corresponding wavelength is smaller than $H$. Another fundamental difference between the two is that, the phase velocity at the pinch point for the confinement-triggered absolute instability follows the Dimotakis speed $U_D$, while the surface-tension absolute instabilities typically exhibit much smaller phase velocities.  
}

{The simulations for the cases with laminar gas inflows (N0 and W0) yield similar results for the dominant  frequency and wavelength, $\omega_L^* = 0.31$ and $\lambda^* = 4.5$. Stability analysis without confinement for the same case has been performed in previous studies \citep{Ling_2019a, Jiang_2020a}, which represent the absolute instability triggered by the surface tension mechanism. The stability analysis slightly overestimates the dominant frequency $\omega^*_\mathrm{st}=0.40$, while the predicted wavelength $\lambda^*_\mathrm{st}=1.6$  is significantly lower than the simulation result, namely the surface-tension absolute instability over-predicted the wavenumber. Furthermore, it is found in the simulation results that, the wave propagation speed $U_w^*=0.22$, which agrees well with the Dimotakis speed and is much higher than the wave speed predicted by the surface-tension absolute instability, $U^*_{w,\mathrm{st}}=0.10$. These discrepancies, together with the fact that $\lambda^*>1$, seem to indicate the absolute shear instability observed here belongs to the inviscid confinement mechanism, instead of the surface-tension mechanism. Nevertheless, to fully confirm the nature of absolute instability, stability analysis similar to \cite{Matas_2015b} needs to be performed, which will be relegated to our future work. }

\subsubsection{Dominant frequency}
As shown in figure \ref{fig:temporal_instability}(b), when the gas inlet turbulence intensity $I$ increases, the dominant frequencies shift to the right for both narrow and wide domains, and the amplitude also increases correspondingly. The normalized frequencies for different cases, $\omega_L^*/\omega_{L,0}^*$, where $\omega_{L,0}^*$ represents the frequency for $I=0$, are plotted in figure \ref{fig:temporal_instability}(c). The experimental data of \cite{Matas_2015a} are also shown for comparison. The experiments covered a wide range of injection conditions for different gas and liquid velocities, but the normalized frequencies approximately collapse. Though the simulation conditions are not identical to those for the experiments, the simulation results for the normalized frequency agree reasonably well with the experimental results. {In particular, the general variation trend of $\omega_L^*/\omega_{L,0}^*$ over $I$ is well captured, \ie, $\omega_L^*/\omega_{L,0}^*$ varies little for $I\lesssim 0.02$ and increases with $I$ for $I\gtrsim 0.02$.} The discrepancy between the simulation and experiment results may be due to the different density and viscosity ratios used in the present simulation. 

The increase of $\omega_L^*$ over $I$ can be explained by the viscous spatial-temporal stability analysis with eddy viscosity model \citep{Naraigh_2013a,Matas_2015a,Jiang_2020a}. The effective gas viscosity is the sum of the molecular and eddy viscosities, which increases with $I$ since the turbulent eddy viscosity increases with $I$. Furthermore, the Orr-Sommerfeld system indicated that the dominant frequency increases with the effective gas viscosity. When $I$ is small, the eddy viscosity is smaller than the molecular counterpart and that is why $\omega_L^*$ varies little for $I\lesssim 0.02$. More discussions on the stability analysis can be found in our previous study \citep{Jiang_2020a}. 

For similar $I$, $\omega_L^*/\omega_{L,0}^*$ for the wide domain is smaller than that for the narrow domain. For example, the values of $I$ for the cases N3 and W1 are similar, $I_{N3}=0.056$ and $I_{W1}=0.06$, respectively, but $\omega_{L,W1}^*=0.38$ is about 9\% lower than $\omega_{L,N3}^*$. This seems to indicate that, the longitudinal instability is more sensitive to domain width when the gas inflow is turbulent. Since the turbulent gas flow is 3D in nature, the small domain {width} may have constrained the development of the turbulence and the turbulence-interface interaction. 
%When the inlet gas is laminar, $\omega_L$ changes little when the domain width varies. Therefore, a narrow domain may be used to predict the dominant frequency, if the inlet gas is laminar. Yet a wide domain is necessary to capture the frequency if the gas inflow is turbulent. 

\subsubsection{Spatial growth}
\begin{figure}
\centering
\includegraphics[trim={0cm 0cm 0cm 0cm},clip,width=.8\textwidth]{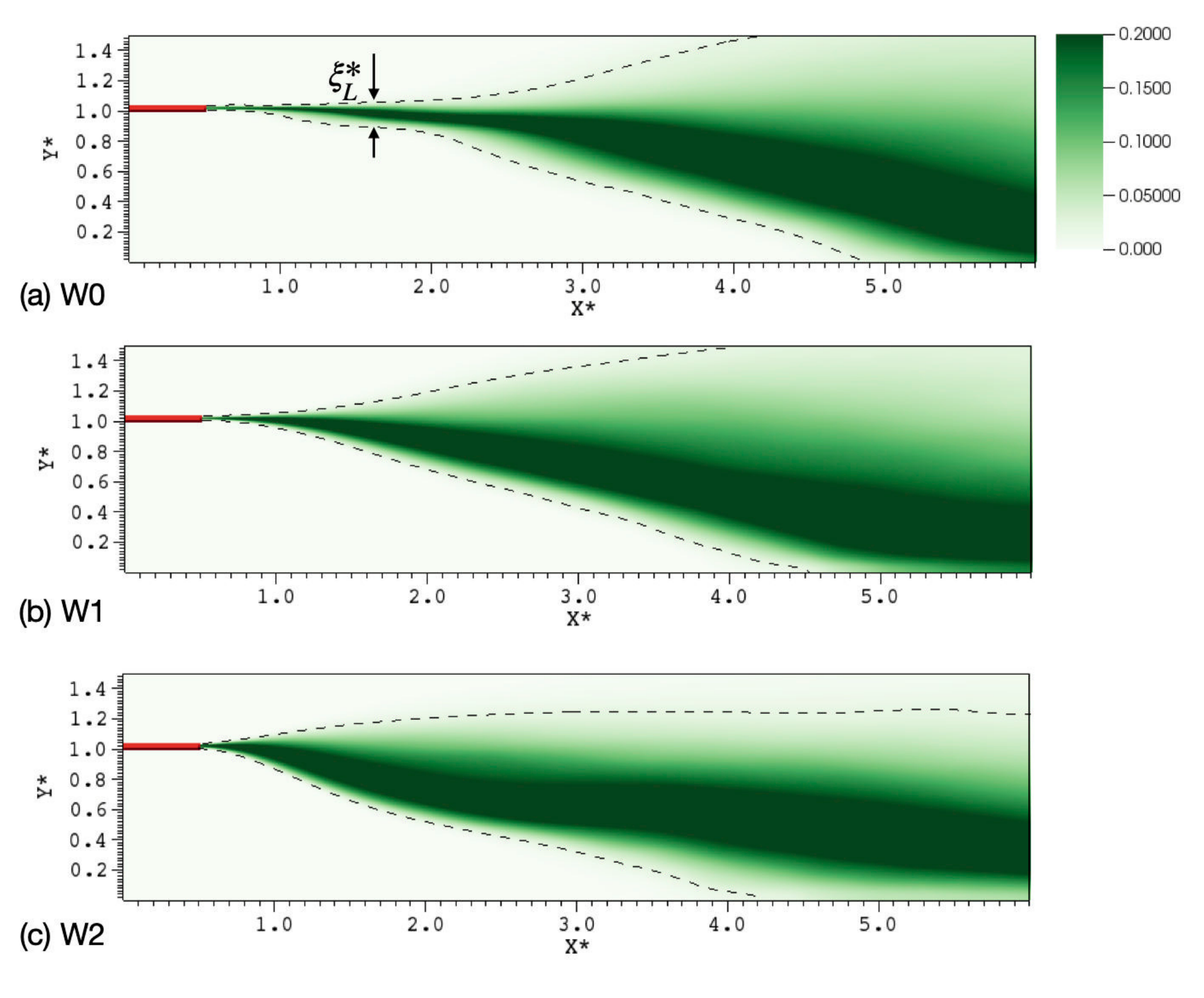}
\caption{Spatial growth of the longitudinal wave amplitude (characterized by the mean square of liquid volume fraction fluctuations $\langle \overline{c'c'}\rangle $, for the cases {(a) W0, (b) W1 and (c) W2}. The contour line $\langle \overline{c'c'} \rangle=0.02$ is used to measured the wave amplitude $\xi_L^*$. }
\label{fig:cc}
\end{figure}

{The spatial development of the longitudinal instability determines the longitudinal growth of the interfacial wave amplitude. Since the interface motion will introduce a fluctuation in the liquid volume fraction $c'=c-\langle \overline{c} \rangle$, the mean square of which, \ie, $\langle \overline{c'c'} \rangle$, is employed to  measure the spatial longitudinal wave amplitude, see figure \ref{fig:cc}. For a given $x^*$, the vertical distance between the two contour lines of $\langle \overline{c'c'} \rangle=0.02$ is defined as the longitudinal wave amplitude $\xi_L^*$. }

\begin{figure}
\centering
\includegraphics[trim={0cm 0 0cm 0},clip,width=.95\textwidth]{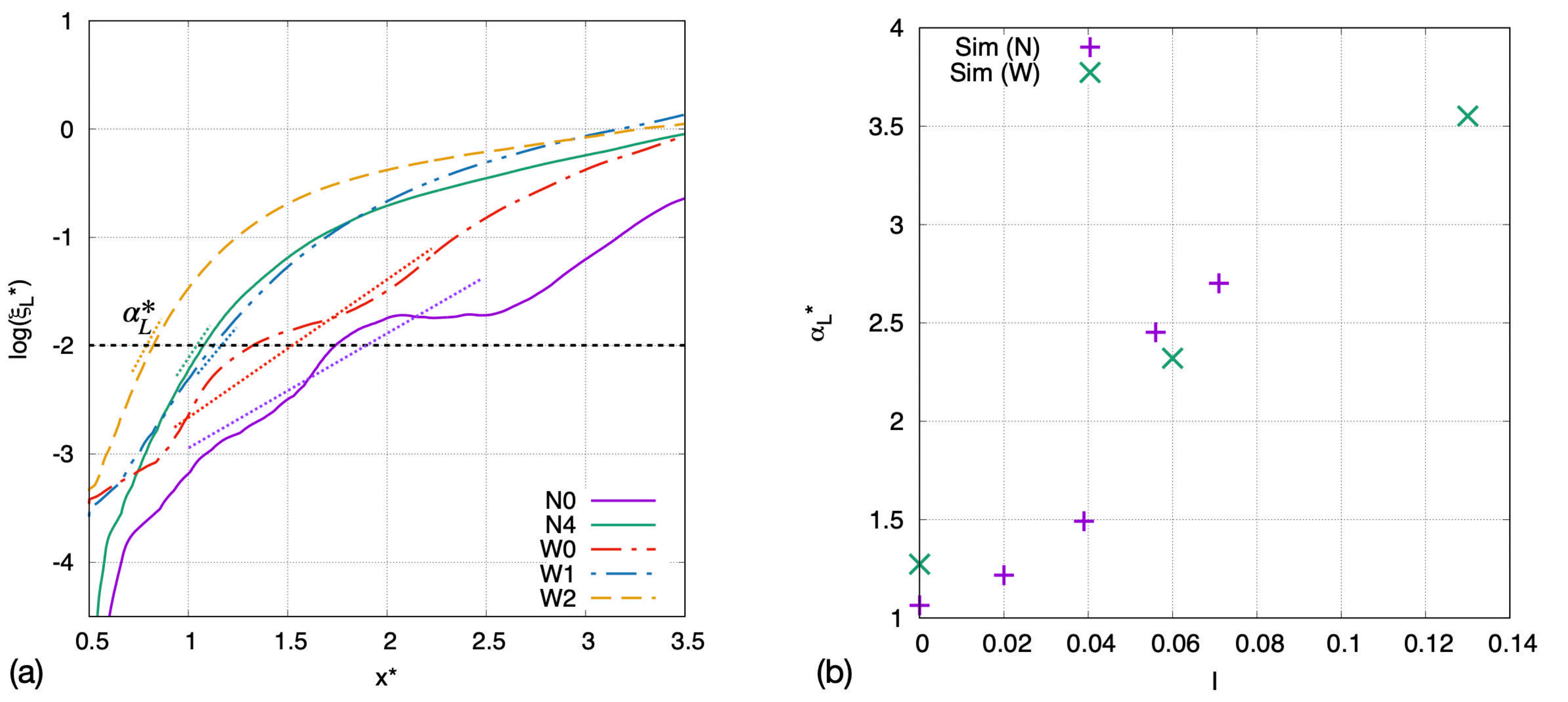}
\caption{(a) The growth of the longitudinal wave amplitude $\xi_L^*$ for different cases. {The dotted lines indicate the spatial growth rate $\alpha^*_L$ for different cases at $\log(\xi_L^*)\approx -2$.} (b) Variation of the normalized spatial growth rate {$\alpha_L/\alpha_{L,0}$} as a function of $I$. } 
\label{fig:growth_rate}
\end{figure}

{It can be seen that, $\xi_L^*$ grows rapidly with $x^*$ near the nozzle exit and becomes more gradually downstream. From the log-linear plot shown in figure \ref{fig:growth_rate}(a), the spatial growth of $\xi_L^*$  near the nozzle exit is approximately exponential, see \eg, $0.5\lesssim x^* \lesssim 0.8$ for the case W2.  Nevertheless, it should be reminded that, since the longitudinal instability here is absolute, nonlinearity will influence the properties of the most-unstable mode and the spatial growth does not follow an exponential function as in convective instability \citep{Otto_2013a,Matas_2015b}. 
}

{As the inlet gas turbulence intensity $I$ increases, the spatial growth of $\xi_L^*$, particularly near the nozzle exit, becomes faster. To characterize the effect of $I$, the spatial growth rate, $\alpha_L^*$, namely the slope of the curve $\log{\xi_L^*}$-$x^*$, is measured at $\log(\xi_L^*)\approx -2$ for all cases. The variation of $\alpha_L^*$ over $I$ is shown in figure \ref{fig:growth_rate}(b). Similar to the dominant frequency, $\alpha_L^*$ also increases with $I$. If a different threshold value for $\langle \overline{c'c'} \rangle$ other than 0.02 is to be used, the values of $\xi_L^*$ would change, but those for $\alpha_L^*$ will not be affected. For the cases W0 and N0, the variation $\xi_L^*$ over $x^*$ exhibits fluctuations, which makes the measurement of $\alpha_L^*$ somewhat sensitive. As will be discussed later in section \ref{sec:transv_amp}, these fluctuations are induced by the spatial averaging over the transverse direction. To reduce the influence of the fluctuations, the measurement of $\alpha_L^*$ for the cases N0 and W0 is made for a wider range of $x^*$, as indicated in figure \ref{fig:growth_rate}(a). }

\begin{figure}
\centering
\includegraphics[trim={0cm 0cm 0cm 0cm},clip,width=.95\textwidth]{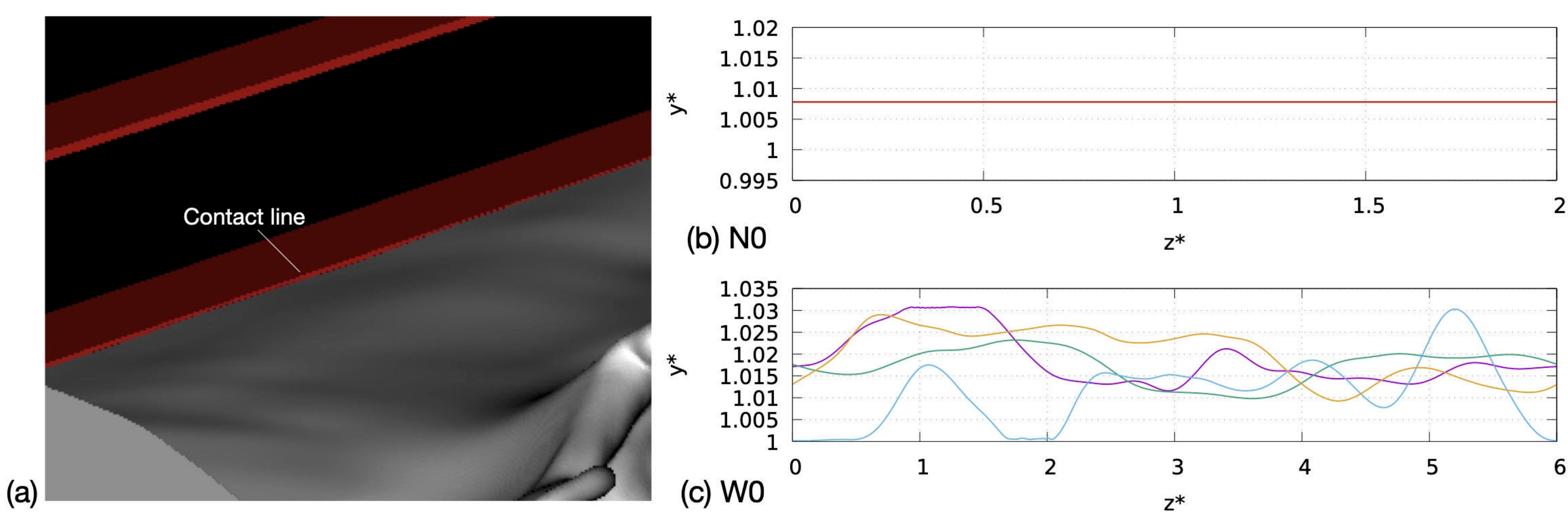}
\caption{(a) Closeup of the interface near the inlet and interfaces at $x^*=\eta_x^*=0.5$ for different times for the cases (a) N0 and (b) W0. The interfaces for the case N0 collapse to a straight line. }
\label{fig:Inlet_cvof}
\end{figure}

It can be observed from \ref{fig:growth_rate}(a) that the values of $\xi_L^*$ at the end of the separator plate, $x^*=\eta_x^*=0.5$, are almost zero for the narrow domain cases, but are finite for the wide domain cases. A snapshot of the interface near the inlet for the case W0 is shown in figure \ref{fig:Inlet_cvof}(a). The contact line where the three phases meet can be seen. A closeup in figure \ref{fig:Inlet_cvof}(c) shows that the contact line varies in time and also in the transverse direction. This indicates that the contact line moves on the separator front wall (facing the streamwise direction). Accurately capturing the contact-line dynamics is out of the scope of the present study. The boundary conditions on the solid separator plate are no-slip wall for velocity and symmetric for the liquid volume fraction $c$ (yielding a 90 degree contact angle). The small-amplitude spatial variation of the contact line is due to a numerical slip, with half of the cell size as the slip length \citep{Snoeijer_2013a}. For the case N0, due to the constraint of the smaller domain width, the contact line remains  to be  a straight line pinned on the lower edge of the separator front wall, see figure \ref{fig:Inlet_cvof}(b). Therefore, when we compute the longitudinal wave amplitude by averaging $\overline{c'c'}$ in the transverse direction, the amplitude at $x^*=\eta_x^*$ for the case N0 is identical to zero, while that for the case W0 is finite. 

Finally, it is worth mentioning that linear stability analysis has been performed in previous studies to identify the most unstable mode in the present configuration. Among the many attempts, the viscous spatial-temporal analysis has been shown to be quite successful \citep{Otto_2013a, Matas_2015b}. The perturbation is introduced in the form of a 2D normal mode. The resulting Orr-Sommerfeld equations are solved to obtain the most-unstable frequency and growth rate, which have been shown to agree well with the experimental and DNS results when the inlet gas is laminar \citep{Fuster_2013a}. In order to predict the most unstable mode when turbulence is present at the gas inlet, additional modeling efforts are required. Attempts have been made by \cite{Matas_2015a} and \cite{Jiang_2020a} by incorporating the effect of inlet turbulence using the turbulent viscosity model. As the turbulent viscosity increases with $I$, the Orr-Sommerfeld equations can reproduce the trends that the frequency increases with $I$. However, the modified stability model underestimates the values. The present simulation results indicate that the interface exhibits 3D features right at the end of the separator plate, the assumption of 2D mode in conventional stability analysis may need to be relaxed to yield better prediction.

\subsection{Transverse instability and wave development}
% RT instability induced by longitudinal-instability-induced vertical interface motion 
\subsubsection{Rayleigh-Taylor instability}
\begin{figure}
\centering
\includegraphics[trim={0cm 0cm 0cm 0cm},clip,width=.65\textwidth]{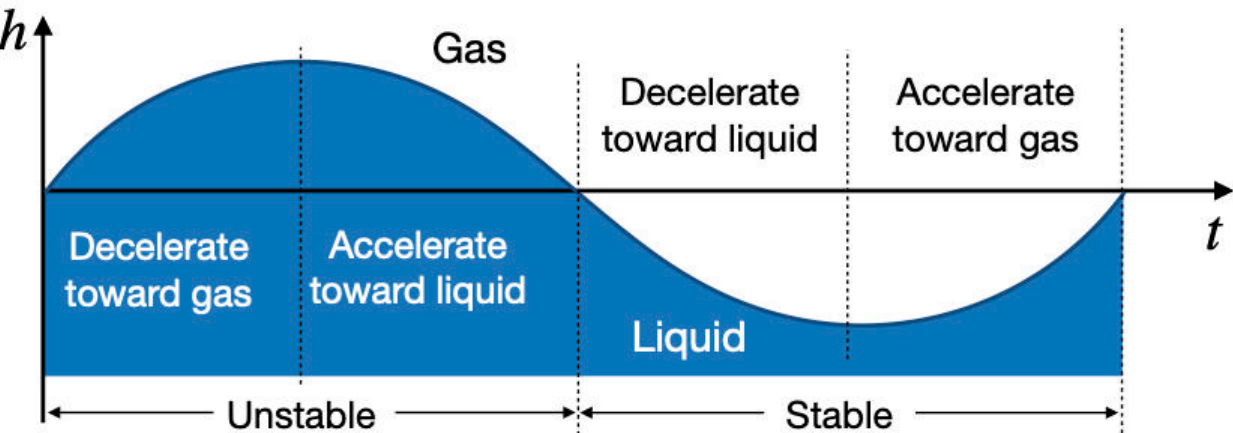}
\caption{Schematic for the transverse Rayleigh-Taylor (RT) instability due to the vertical motion of interface induced by {the} longitudinal instability.}
\label{fig:RT_instability}
\end{figure}

Though the interfacial waves near the inlet are approximately longitudinal, transverse modulations are also observed, see figure \ref{fig:general_behavior}. The transverse modulations are induced by the Rayleigh-Taylor (RT) instability. As shown in figure \ref{fig:temporal_instability}, the longitudinal instability introduces oscillatory motion of the interface in the vertical direction. When the interface accelerates toward the liquid or decelerates toward the gas, the interface is unstable due to the baroclinic effect and the RT instability is triggered, see figure \ref{fig:RT_instability}. While the longitudinal instability develops right at the end of the separator plate, the transverse interfacial modulation may not grow immediately, see figure \ref{fig:general_behavior}(a). The reason is that the interface is stable in half time of one oscillation cycle (when the interface decelerates toward the liquid or accelerates toward the gas), {see figure \ref{fig:RT_instability}}, so the transverse instability may first grow and then decay. 

\begin{figure}
\centering
\includegraphics[trim={0cm 0cm 0cm 0cm},clip,width=.95\textwidth]{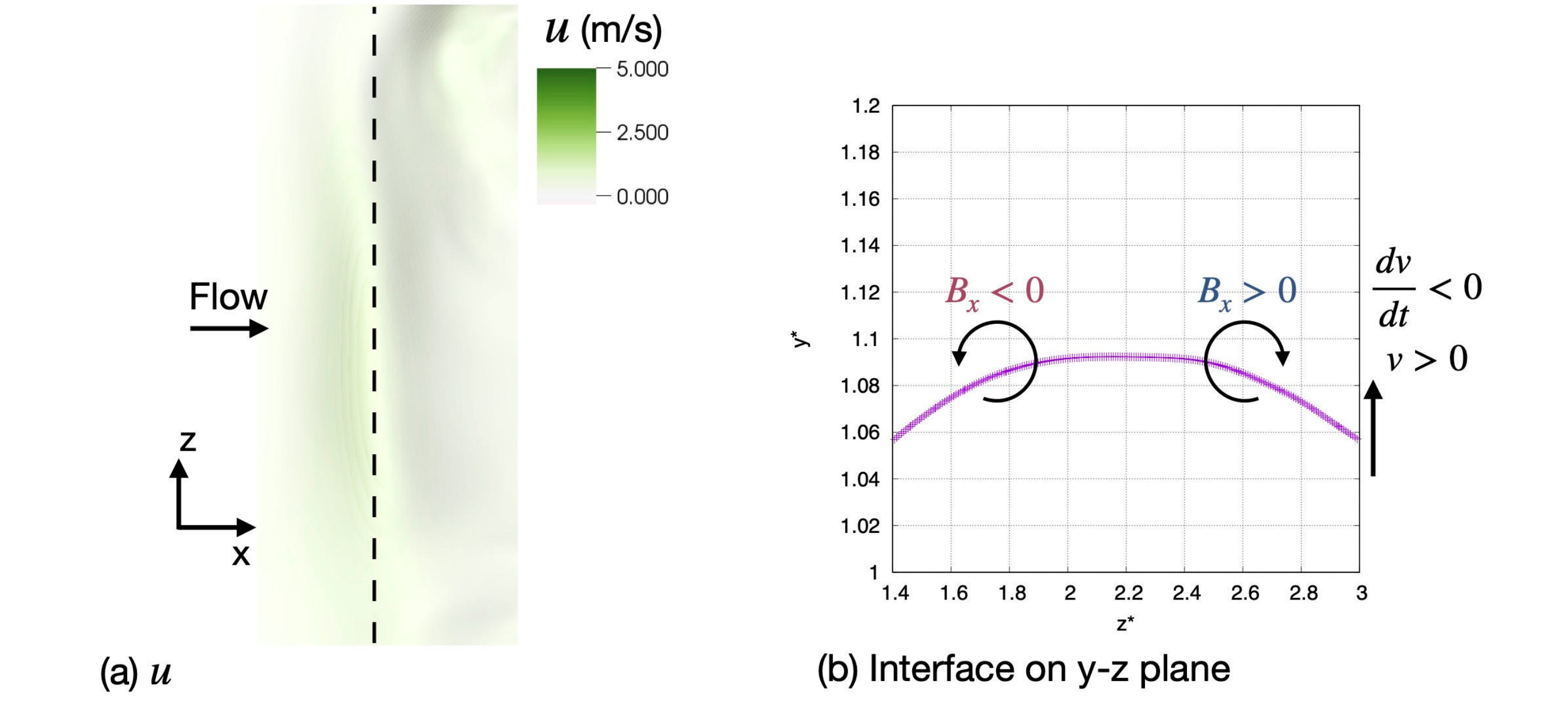}
\caption{Transverse development of the Rayleigh-Taylor instability for the case W0. (a) Top view of the interface at $t^{*} = 323.75$, and the interface is colored with $u$-velocity $(m/s)$. {The vertical dashed line in (a) indicates the wave location, and the interface profile along the dashed line is shown in (b) on the $y$-$z$ plane to demonstrate the baroclinic effect.}}
\label{fig:transverse_instability}
\end{figure}

% RT instability plus Bernoulli depression to cause the 3D wave
The present results indicate that two conditions need {to} be satisfied for a transverse modulation to grow and to transform the 2D longitudinal wave to fully 3D. First, the longitudinal wave amplitude must be sufficiently large. Second, the interface is decelerating toward the gas, so that the growing interfacial wave will interact with the gas stream. A representative example of the development of the transverse instability is shown in figure \ref{fig:transverse_instability}. The later time evolution of this specific wave can be found in figure {\ref{fig:general_behavior}(c)}. At this time ($t^*=323.75$) the interfacial wave is well aligned with transverse direction if viewed from the top, with some small transverse variations in the height of the wave crest.  The interface is rising upward and decelerating toward the gas, the misalignment between the pressure and density gradients generates a baroclinic torque. The baroclinic torque in $x$ direction, $B_x= (\nabla \rho \times \nabla p)_x$ will cause the transverse interfacial perturbation to grow, as indicated in figure \ref{fig:transverse_instability}(b), which shows the interface profile on the $y$-$z$ plane. At this streamwise location $x^*\approx 1.6$, the amplitude of longitudinal wave is large {enough}, as a result, the wave crest will experience a significant interaction with fast gas stream above the wave. The local acceleration of the gas flow above the wave crest creates a Bernoulli depression, which will further pull the interface upward \citep{Hoepffner_2011a}, amplifying the transverse modulation of the interface, see $t^*=322.5$ to 327.5 in figure {\ref{fig:general_behavior}(c)}. More discussions about the interaction between the interfacial wave and the gas stream will be presented in section \ref{sec:wave-gas_interact}. 

\subsubsection{Dominant mode for transverse instability}
Though the transverse modulation growth is assisted by its interaction with the gas stream, the selection of the dominant transverse wavenumber is mainly controlled by the RT instability. The inviscid linear temporal RT instability with surface tension for a planar interface yields the following dispersion relation \citep{Chandrasekhar_1961a}
\begin{align}
	\omega_I = \left(ak\frac{\rho_l-\rho_g}{\rho_l+\rho_g}- \frac{\sigma k^3}{\rho_l+\rho_g}\right)^{1/2}\,,
	\label{eq:RT_dispersion}
\end{align}
where $\omega_I$ is temporal growth rate, $k$ is wavenumber, and $a$ is magnitude of the interfacial acceleration. To account for the viscous effect, the dispersion relation becomes \citep{Chandrasekhar_1961a, Joseph_1999a}
\begin{align}
	\omega_I = -k^2 \frac{\mu_l+\mu_g}{\rho_l+\rho_g}\pm \left(ak\frac{\rho_l-\rho_g}{\rho_l+\rho_g}- \frac{\sigma k^3}{\rho_l+\rho_g}+ k^4 \left(\frac{\mu_l+\mu_g}{\rho_l+\rho_g}\right)^2 \right)^{1/2}\,.
	\label{eq:RT_dispersion_visc}
\end{align}
Due to the low liquid viscosity in the present problem, the viscous effect on the RT instability is very small. The difference between Eqs.\ \eqr{RT_dispersion} and \eqr{RT_dispersion_visc} is negligible, so the inviscid relation Eq.\ \eqr{RT_dispersion} will be used in the following analysis. 

Based on Eq.\ \eqr{RT_dispersion}, the transverse wavenumber for the most unstable mode is 
\begin{align}
	k_{T}=\sqrt{\frac{a(\rho_l-\rho_g)}{3\sigma}}\, ,
	\label{eq:k_RT_1}
\end{align}
and the corresponding growth rate is 
\begin{align}
	\omega_{T}=\frac{2}{3^{3/2}} \left( \frac{a^3(\rho_l-\rho_g)}{\sigma}\right)^{1/4}\, .
	\label{eq:omega_RT_1}
\end{align}

The interfacial acceleration $a$ is induced by the longitudinal instability, for which the characteristic length and time scales are $\delta$ and $1/\omega _L$, respectively. Therefore, it can be approximated that $a\sim \delta \omega_L^2$ and the dominant transverse wavenumber 
\begin{align}
	k_{T}\approx \omega_L\sqrt{\frac{\delta  (\rho_l-\rho_g)}{3\sigma}}\, .
	\label{eq:k_RT}
\end{align}
As shown in section \ref{sec:long_instability}, $\omega_L$ increases with $I$. According to Eq.~\eqr{k_RT}, $k_T$ will also grow with $I$. 

Using similar scaling relations, the temporal growth rate for the transverse instability can be estimated as 
\begin{align}
	\omega_{T} \approx \frac{2}{3^{3/2}} \left( \frac{\delta^3 \omega_L^6(\rho_l-\rho_g)}{\sigma}\right)^{1/4}\, .
	\label{eq:omega_RT_2}
\end{align}
Since $\omega_L \sim U_D/\delta$, it can be shown that 
\begin{align}
	{\omega_{T}} \approx {\omega_L}\text{We}_{L}^{1/4}\, ,
	\label{eq:omega_RT}
\end{align}
where $\text{We}_L$ is the Weber number for the longitudinal wave, expressed as 
\begin{align}
	\text{We}_L = \frac{(\rho_l-\rho_g)U_D^2 \delta}{\sigma}\, . 
	\label{eq:We_L}
\end{align}
If ${\omega_{T}}$ is significantly lower than ${\omega_L}$, then the RT instability will not grow fast enough to be relevant. For the present problem, $\text{We}_L=9.45$, {and} it is estimated that ${\omega_{T}}/{\omega_L} \sim $ O(1). Therefore, the time scales for the longitudinal KH and the transverse RT instabilities are comparable. In other words, the transverse RT instability can grow in a time scale that is comparable to oscillatory interface motion induced  by the longitudinal instability. This affirms that RT instability is responsible to the transverse development of the interfacial waves for the present problem. 
%Furthermore, since $\omega_L$ increases with $I$, the growth rate of the transverse modulation $\omega_T$ will also increase with $I$. 

% Effect of inlet gas turbulence
\subsubsection{Transverse wavenumber spectra}
\begin{figure}
\centering
\includegraphics[trim={0cm 0cm 0cm 1cm},clip,width=1.00\textwidth]{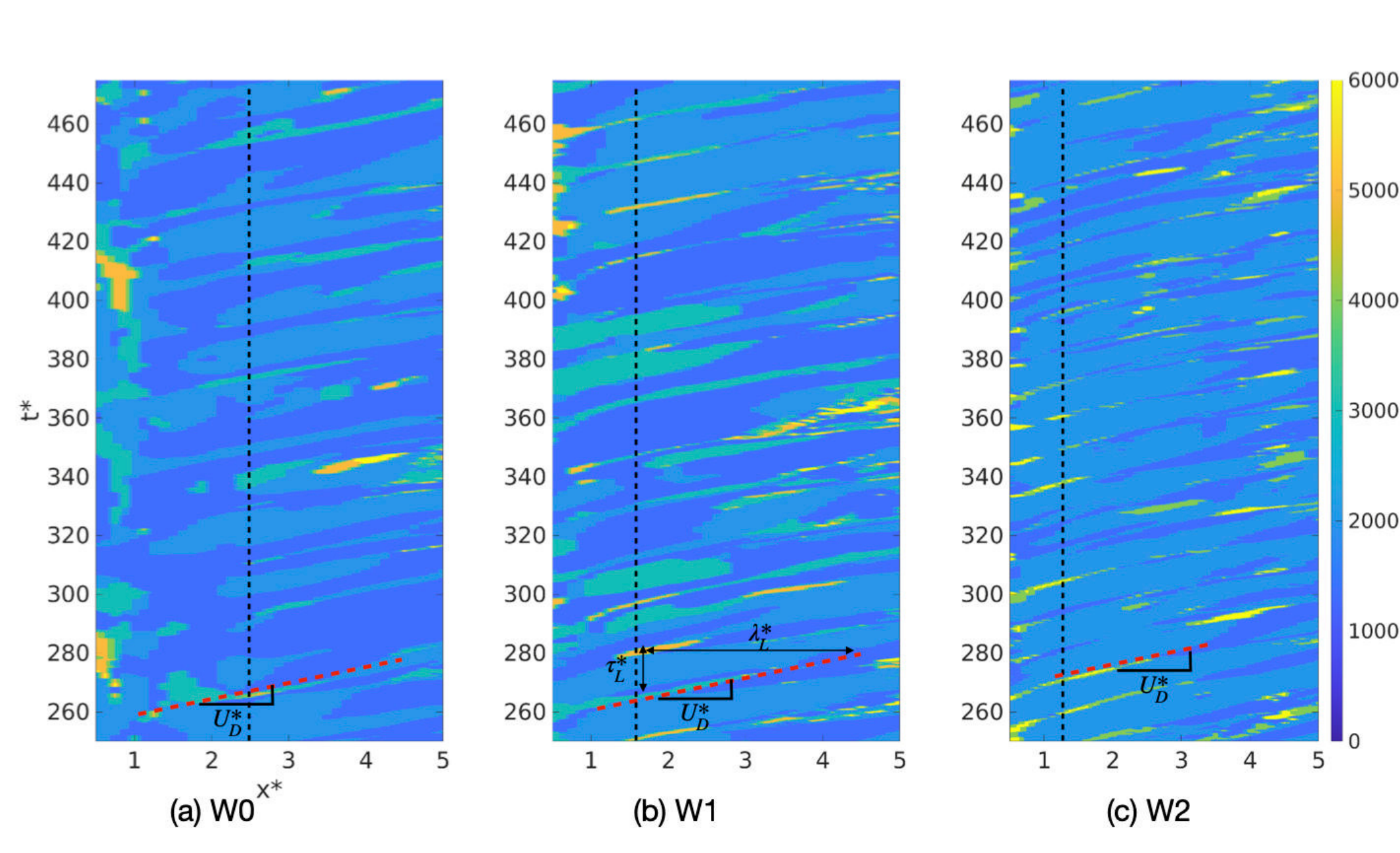}
\caption{The dominant transverse wavenumber $k_{\max}$ on the $x^{*}$-$t^{*}$ diagrams for {the} cases {(a) W0, (b) W1 and (c) W2}. The vertical dashed lines indicate the streamwise locations to estimate time-averaged transverse wavenumber $\overline{k}_T$.}
\label{fig:k_max}
\end{figure}

To investigate the effect of inlet gas turbulence intensity $I$ on the dominant transverse wavenumber, the wavenumber spectra of the transverse interfacial modulation is examined. For a given $t^*$ and $x^*$,  Fourier transform is performed for the {interfacial} height $h$ along the transverse direction. Then the wavenumber for the maximum amplitude in the spectrum, $k_{\max}$, is measured. The contours of $k_{\max}$ for different $I$ are shown in figure \ref{fig:k_max}. 

%k_max rises at longitudinal wave 
It is observed that $k_{\max}$ rises up to much larger values when longitudinal waves pass by. This suggests that the development of the transverse modulations are closely associated with the longitudinal waves, as predicted by the stability analysis (Eq.\ \eqr{k_RT}). As a result, the $k_{\max}$ contours also reveal important features about the formation and propagation in the longitudinal waves. First of all, the trajectories of the longitudinal waves appear as inclined straight lines of higher $k_{\max}$ values in the figures. The slopes of the lines represent the wave propagation speeds in the longitudinal direction. It is observed that the wave speeds are very similar for all the cases and agree with the Dimotakis speed \citep{Dimotakis_1986a}, $U_D$. For the present problem, $U_D^*=U_D/U_g=0.223$. As the longitudinal wave grows in amplitude and interact with the fast gas stream, the aerodynamic drag will cause the wave to accelerate. That is why the longitudinal wave trajectories bend downward slightly further downstream, which is most profound for the case W2 due to the {faster spatial growth} for the longitudinal instability, see figure \ref{fig:k_max}(c). The time period $\tau_L$ and the wavelength $\lambda_L$ for the longitudinal waves can also be identified, and they are related by $U_D$ as $\lambda_L =U_D \tau_L$. It is also observed that $\tau_L$ decreases with $I$. Since $U_D$ remains unchanged, $\lambda_L$ decreases with $I$. 

\begin{figure}
\centering
\includegraphics[trim={0cm 0cm 0cm 0cm},clip,width=.95\textwidth]{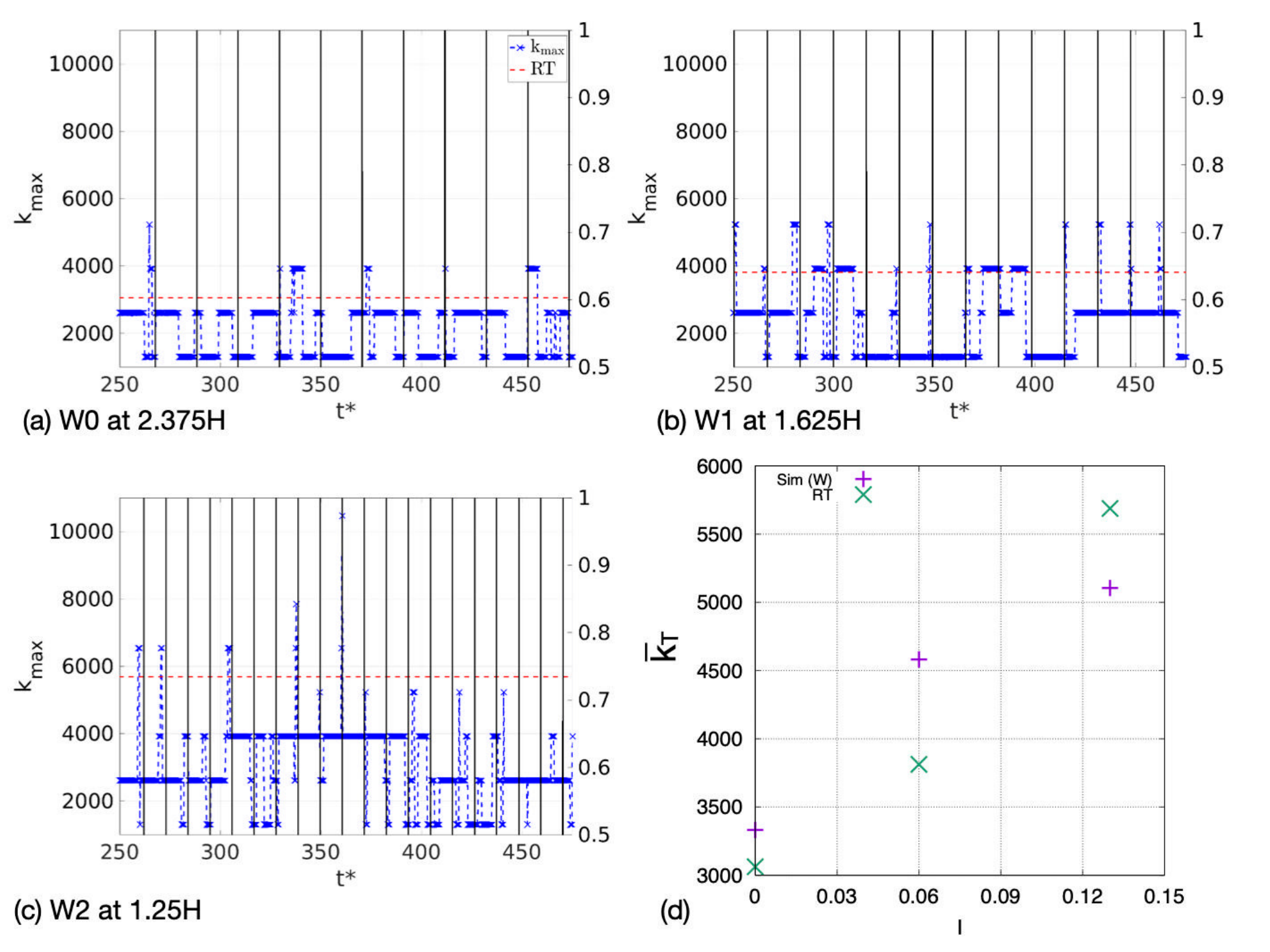}
\caption{Temporal evolution of transverse wavenumber $k_{\max}$ for the cases (a) W0, (b) W1, and (c) W2, respectively. The vertical lines indicated the passage times of the longitudinal waves, which are estimated based on the $\omega_L$for the corresponding $I$. (d) Variation of $\overline{k}_{T}$ with turbulent intensity $I$, compared with the RT predictions (Eq.\ \eqr{k_RT}).}
\label{fig:k_vs_I}
\end{figure}

%Increase of k_T over I
It is noted that only $k_{\max}$ on the longitudinal wave location characterizes the transverse modulations of the interfacial waves. Between two longitudinal waves, $k_{\max}$ represents the ripples on the interfaces, which are less important. Therefore, we denote the $k_{\max}$ at the passages of the longitudinal waves as the transverse wavenumber of the interfacial wave, $k_T$. It is observed that $k_{T}$ for different waves are different due to the chaotic nature of the turbulent multiphase flows. The temporal evolutions of $k_{max}$ for different $I$ is shown in figures \ref{fig:k_vs_I}(a)-(c). For each case, the measurement is made at $x^*$ locations corresponding to $\log(\xi_L^*)=-1$, \ie, $x^*=2.375$, 1.625, 1.25, for the cases W0, W1, and W2, respectively, which are indicated by dashed lines in figure \ref{fig:k_max}. The RT instability predictions (Eq.~\eqr{k_RT}) are also shown in figure \ref{fig:k_vs_I} for comparison. It can be observed that $k_{T}$ generally increases when $I$ increases, which is consistent with the RT instability theory. The increase of $k_T$ with $I$ is also consistent with the snapshots shown in figure \ref{fig:general_behavior}. 

The time average transverse wavenumber $\overline{k}_{T}$ is plotted as a function of $I$ in figure \ref{fig:k_vs_I}(d). The measurement times for  $k_{T}$ are indicated by the vertical lines in figures \ref{fig:k_vs_I}(a)-(c), which correspond to the longitudinal wave passage times estimated based on the dominant longitudinal frequency $\omega^*_L$ for the given $I$. It is shown that $\overline{k}_T$ increases from about 3330 to 5100 when $I$ increases from 0 to 0.13. The values of $\overline{k}_T$ for $I=0$ and 0.13 correspond to about 2.5 and 3.9 waves in the domain width. {The RT instability model (Eq.~\eqr{k_RT}) well captures the increasing of trend of $\overline{k}_T$ over $I$. The discrepancy between the model predictions and the simulation results can be up to 17\%, so the model predictions can only be used as approximations of $\overline{k}_T$.}

\subsubsection{Transverse modulation amplitude}
\label{sec:transv_amp}
\begin{figure}
\centering
\includegraphics[trim={0cm 15cm 0cm 0},clip,width=.95\textwidth]{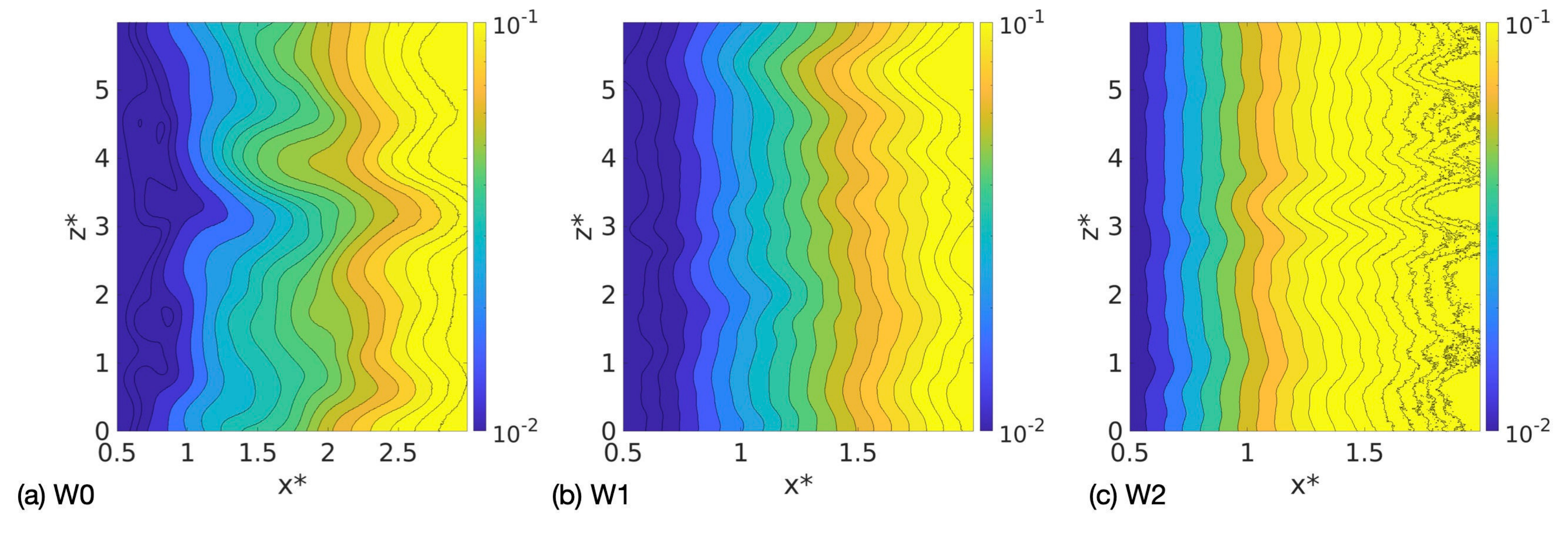}
\caption{Spatial variation of the interfacial wave amplitude $\xi^*=(\overline{h'h'})^{1/2}/H$ for {the cases (a) W0, (b) W1, and (c) W2.}}
\label{fig:hh}
\end{figure}

%Increase of amplitude over I
The wave amplitude for a given $(x^*,z^*)$ location can be characterized by the root mean square of the interfacial height fluctuations normalized by $H$, \ie, $\xi^*=(\overline{h'h'})^{1/2}/H$. The variations of $\xi^*$ with $x^*$ and $z^*$ for different cases are shown in figure \ref{fig:hh}. The wave amplitude generally increases in the longitudinal direction, though small variations in the transverse direction are also observed. It should be noted that $\xi^*$ can be used to characterize the wave amplitude only when the wave amplitude is small. The noises observed in figure \ref{fig:hh}(c) are due to the rolling up and breakup of the waves. The transverse variations are the most profound for the case W0. 

\begin{figure}
\centering
\includegraphics[trim={0cm 0cm 0cm 0},clip,width=.95\textwidth]{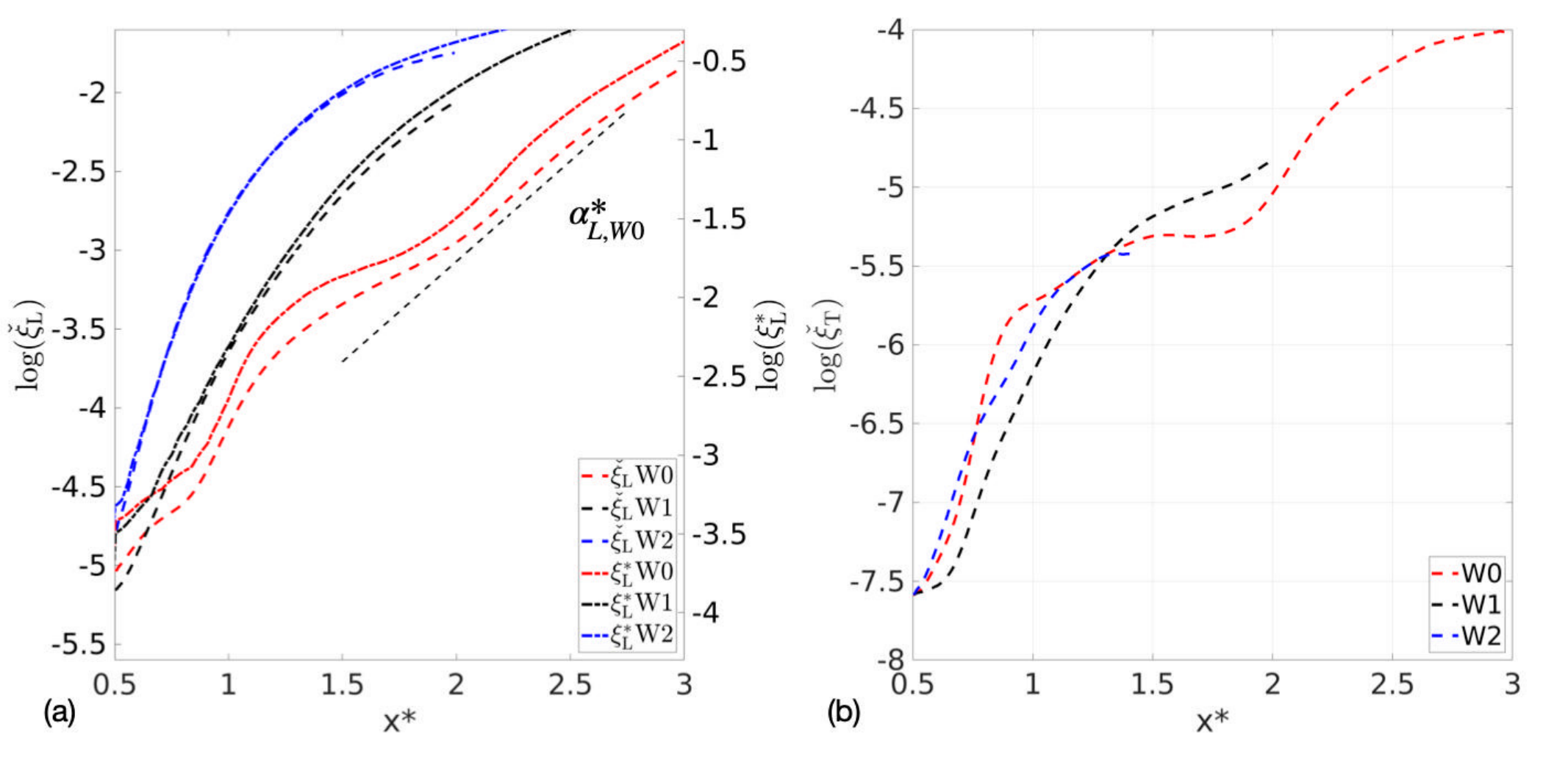}
\caption{Spatial variation of (a) longitudinal wave amplitude $\check{\xi}_L$ and (b) transverse modulation amplitude $\check{\xi}_T$ along the $x^*$ direction for the cases W0, W1 and W2. The results for the longitudinal wave amplitude $\xi_L^*$, {which are} measured based on the liquid volume fraction fluctuations $\langle \overline{c'c'}\rangle$, are also shown in (a) for comparison.}
\label{fig:hh_line}
\end{figure}

\begin{figure}
\centering
\includegraphics[trim={0cm 0cm 0cm 0},clip,width=.95\textwidth]{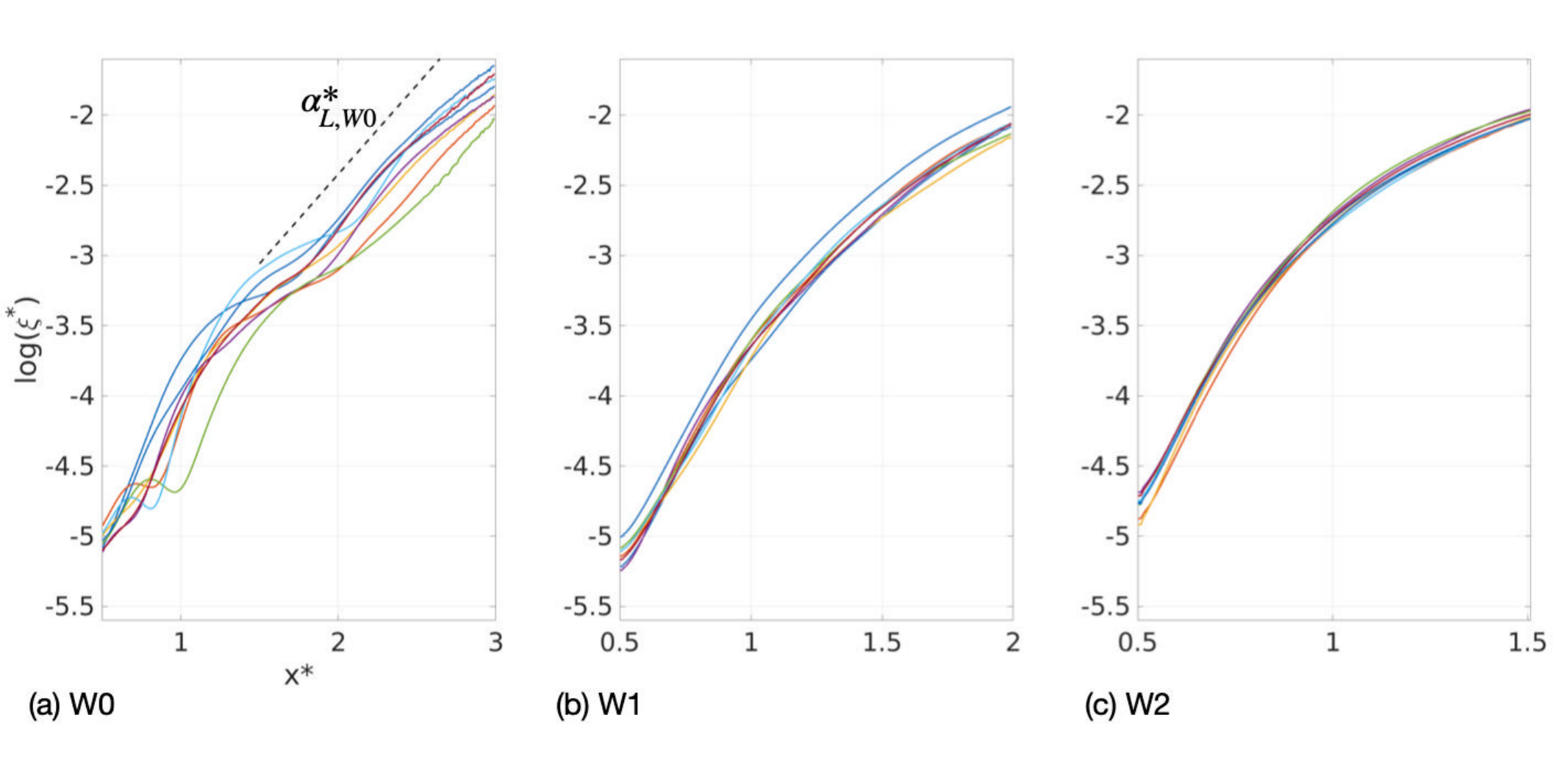}
\caption{Spatial variation of wave amplitude $\xi^*$ at different $z$ locations {for the cases (a) W0, (b) W1 and (c) W2.}}
\label{fig:hh_variation}
\end{figure}

If $\xi^*$ is averaged in the transverse direction, we can obtain the longitudinal wave amplitude, \ie, $\check{\xi}_L = \langle (\overline{h'h'})^{1/2}/H\rangle$. The spatial variation of $\check{\xi}_L$ in $x^*$ direction is shown in figure \ref{fig:hh_line}(a). The results for the longitudinal wave amplitude $\xi_L^*$ defined based on the contours of $\langle \overline{c'c'}\rangle=0.02$ (see figure \ref{fig:cc}(a)) are also shown for comparison. It can be observed that though the values for $\check{\xi}_L$ and ${\xi}_L^*$ are different, the ratio $\check{\xi}_L/{\xi}_L^*$ is approximately a constant for all cases. Therefore, the spatial growth rates for the longitudinal instability $\alpha_L^*$ will remain the same, no matter $\check{\xi}_L$ or ${\xi}_L^*$ are used. The curves of $\check{\xi}_L$ for the case W0 exhibits some fluctuations, which are due to stronger transverse variations of $\xi^*$. The results of $\xi^*$ at different transverse locations are shown in figure \ref{fig:hh_variation}. It can be observed that the $\xi^*$-$x^*$ curves are very different for the case W0, while the difference among different profiles is much smaller for the cases W1 and W2. In spite of the different $\xi^*$-$x^*$ curves at different transverse locations for the case W0, the growth rates $\alpha_{L,W0}^*$ are quite similar, see figure \ref{fig:hh_variation}(a). Therefore, the measurement of $\alpha_L^*$ for the case W0 is not as uncertain as it may appear in figures \ref{fig:growth_rate}(a) and \ref{fig:hh_line}(a).

For a given $x^*$, the deviation of wave amplitude $\xi^*$ from the transverse mean $\check{\xi}_L$ can be calculated as $\xi^*-\check{\xi}_L=(\overline{h'h'})^{1/2}/H-\langle (\overline{h'h'})^{1/2}/H \rangle$. Then the root mean square of the deviation, $\check{\xi}_T=\langle (\xi^*-\check{\xi}_L)^2\rangle^{1/2} $, can be used to characterize the average amplitude of the transverse modulations, which is plotted as a function of $x^*$ in figure \ref{fig:hh_line}(b). Near the inlet, $\check{\xi}_T$ is at least an order of magnitude smaller than $\check{\xi}_L$ ($\check{\xi}_L/\check{\xi}_T \gtrsim \exp(2.5)$). As the longitudinal wave amplitude grows in $x^*$, the transverse modulation amplitude also grows correspondingly. The spatial growth rates of $\check{\xi}_T$ near the inlet are similar for different $I$. Yet since the spatial growth of  $\check{\xi}_L$ is higher for larger $I$, the wave amplitude exceeds the linear regime at a smaller $x^*$. Beyond that point, the interfacial waves start to deform and break, {then} $\check{\xi}_T$ is not valid to characterize the transverse modulation amplitude. For the case W0, since the longitudinal wave grows more slowly, $\check{\xi}_T$ has a wider $x^*$ range to grow, reaching a higher value than for the cases W1 and W2. 

\subsection{Development and disintegration of 3D interfacial waves}
\subsubsection{Wave-gas-stream interaction}
\label{sec:wave-gas_interact}
\begin{figure}
\centering
\includegraphics[trim={0cm 0 0cm 0},clip,width=1\textwidth]{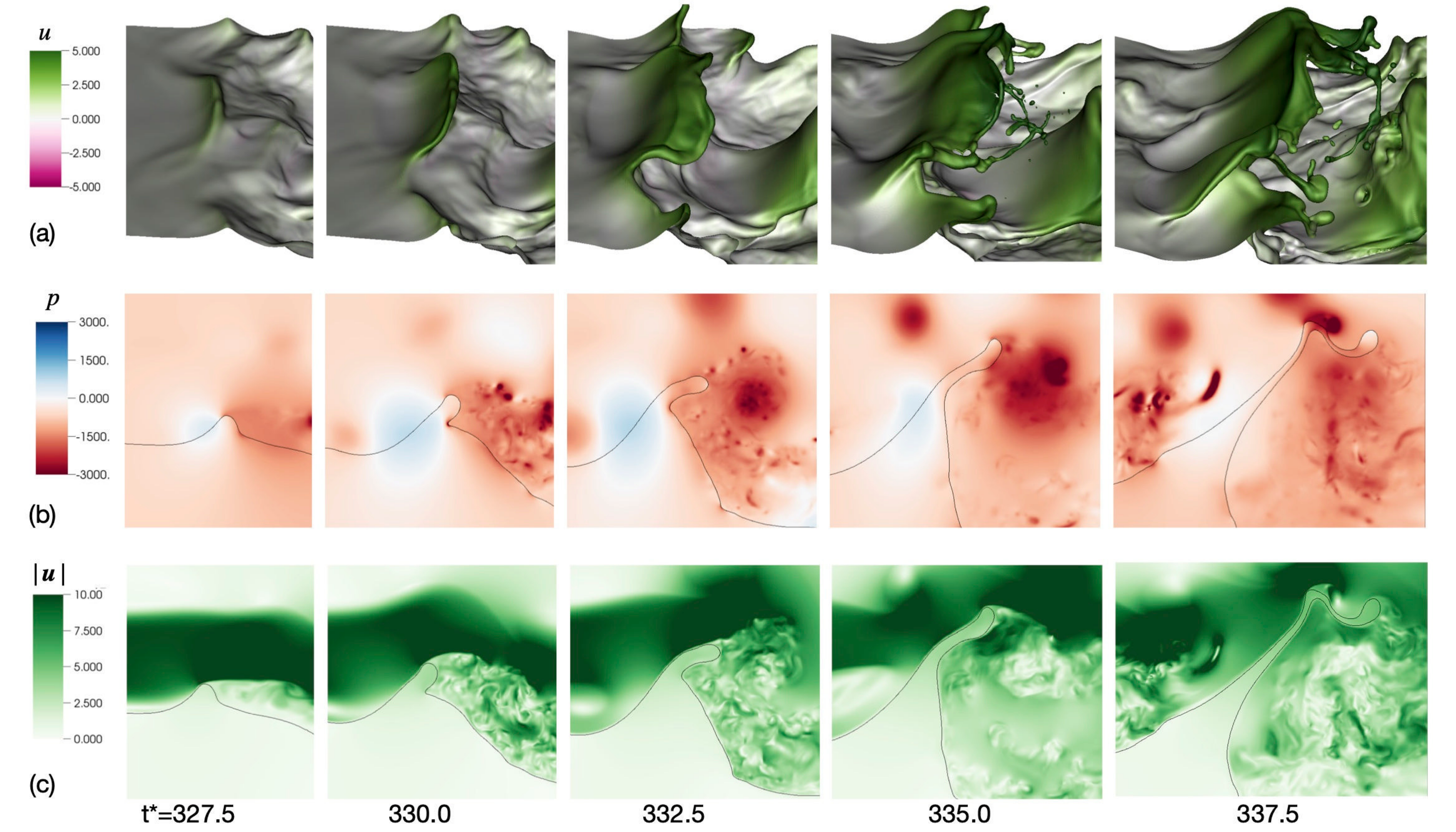}
\caption{Temporal evolutions of the (a) interfaces, (b) pressure, and (c) velocity magnitudes for the case W0, to demonstrate the interaction between the interfacial wave and the gas stream. The interfaces in (a) are colored by the streamwise velocity. The pressure and velocity shown in (b) and (c) are on the $x-y$ plane through the wave. {The frame is moving with the wave to keep the wave located at the center of the window.}}
\label{fig:u_surf}
\end{figure}

%(XXX Wave-gas interaction and wave deformation, showing streamwise velocity on interface. Different behaviors for $I=0$ and 0.13. For $I=0$, wave acceleration and Rayleigh-Taylor instability, contributing to transverse variation of the wave front/rim. For $I=0.13$, merge of waves? Rayleigh-Plateau instability and finger formation? XXX) 
When the wave amplitude grows beyond the linear regime, such as $\xi^*>0.1$, the non-linear interaction between the interfacial wave and the fast gas stream will influence the subsequent deformation of the wave and the liquid sheet extended from the wave crest. The temporal evolutions of the interfaces, the pressure, and the velocity magnitude for the case W0 are shown in figure \ref{fig:u_surf}. The first snapshots at $t^*=327.5$ correspond to approximately the end of the linear regime, and the interfacial wave appears like a lobe of small amplitude. The continuous growth of the wave introduces perpendicular exposure of the wave crest to the fast gas stream. As a consequence of that, a high pressure is built up on the upstream side of the wave, see figure \ref{fig:u_surf}(b). The gas flow accelerates over the wave crest, introducing a low pressure above due to the Bernoulli principle. Furthermore, the gas flow separates on the downstream side (see figure \ref{fig:u_surf}(c)), which also introduces a low pressure wake region.  The pressure difference leads to an aerodynamic drag that pushes the wave crest upward and also along the streamwise direction, further destabilizing the interfacial wave. As a consequence of that, a liquid sheet is extended from the wave crest, see figure \ref{fig:u_surf}(a) at $t^*=332.5$. The Taylor-Culick rim is formed on the edge of the liquid sheet. It can be observed the rim is not as smooth as that in the earlier time and exhibits transverse variation with a higher wavenumber. Multiple mechanisms contribute to the secondary transverse variations of the wave shape. At first, another Rayleigh-Taylor instability is triggered due to the acceleration of the rim in the longitudinal direction. Furthermore, the Rayleigh-Plateau instability of the rim can also lead to variation in the rim diameter. Finally, the turbulent wake of the wave and the interaction between the turbulent vortices and the liquid sheet can also cause irregular deformation of the liquid sheet.

As the interfacial wave continues to grow, the interaction with the gas stream becomes more intense and complex. The liquid sheet starts to bend and fold. The thickness of the liquid sheet at the folding region becomes very small and as a result, holes are formed in the liquid sheet. In figure \ref{fig:u_surf}(a), it can be observed that at $t^*=337.5$ the folded segment of the liquid sheet has completely broken, forming droplets and filaments of different sizes. A new rim will form at the edge of the unbroken sheet that remains attached to the wave, and another cycle of wave-gas-stream interaction will occur. Yet as the interfacial wave is now further away from the inlet and the velocity of the gas stream decreases along the streamwise direction due to turbulent dissipation, the subsequent interaction between the interfacial wave and the gas stream will be less intense. 

\subsubsection{Holes formation and sheet disintegration}
\begin{figure}
\centering
\includegraphics[trim={0cm 0cm 0cm 0cm},clip,width=1.\textwidth]{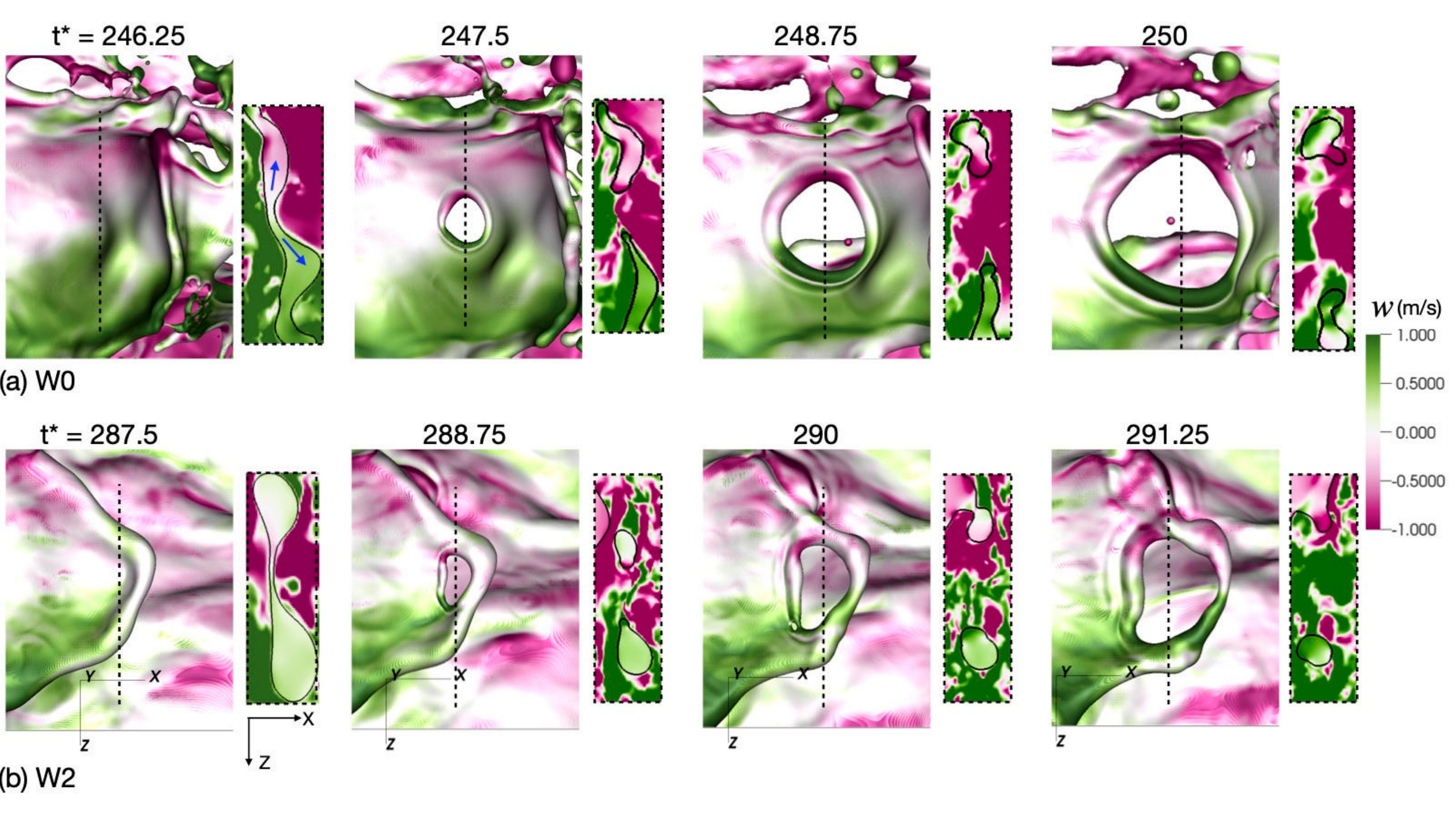}
\caption{Formation and development of holes in liquid sheets for the cases (a) W0 and (b) W2. The $w$-velocity is plotted on the interfaces and also on the cross-section $y$-$z$ planes at the hole centers (indicated by the dashed lines). {The frame is moving with the hole to keep the hole located at the center of the window.}}
\label{fig:holes}
\end{figure}

%(XXX Flapping of liquid sheet. Formation of holes and fingers. Taylor-Culick theory for hole expansion. Measure hole expansion speed in simulations for one specific case, and quantitatively compare the numerical result for hole expansion velocity with TC theory. What is the effect of $I$ in hole formation and expansion? What is the point for figure~\ref{fig:hole_formation}(a) and (b)? XXX)
The formation and development of the holes in the liquid sheet for the cases W0 and W2 are shown in figure \ref{fig:holes}. The $w$-velocity is shown on the interfaces and also on the cross-section $y$-$z$ planes at the hole centers. As can be seen in figure \ref{fig:holes}(a) at $t^*=246.25$, the thickness of the liquid sheet is highly uneven due to the sheet deformation. Furthermore, it can be observed that the liquid is moving away from location of minimum thickness (see figure \ref{fig:holes}(a) at $t^*=246.25$). Eventually the two surfaces pinch and a hole is formed. The pinching process observed here is reminiscent of pinching of liquid neck in drop formation \cite{Castrejon-Pita_2015a, Zhang_2019b}. After the holes are formed, the Taylor-Culick (TC) rim develops on the edge of the hole. Due to the capillary effect, the rim retracts radially, {making the holes expand}. The speed of hole expansion is dictated by the rim retraction velocity, namely the TC velocity, 
\begin{equation}
U_\text{TC}=\sqrt{\frac{2\sigma}{\rho e}}
\label{eq:vel_TC}
\end{equation}
where $e$ is the liquid sheet thickness. The curvature of the rim has little effect on the retraction velocity \citep{Agbaglah_2021a}. Since the thickness of the liquid sheet is non-uniform, $e$ varies from $14.2$  to $6.5$ \textmu m near the edge of the hole, see figure \ref{fig:holes}(a) at $t^*=247.5$, for which the hole expansion velocity predicted by theory (Eq.\ \eqr{vel_TC}) is $0.84-1.24$ m/s. The hole expansion velocity measured from the simulation results is about 1.35m/s, which is in reasonable agreement with the TC predictions. When the inlet gas is turbulent, the gas flow around the sheet exhibits more intense fluctuations, which can be recognized from the footprints on the interfacial velocity (see figure \ref{fig:holes}(b)). The surrounding turbulence seems to have little effect on the hole expansion speed, which is found to be similar to that for the case W0. This indicates that the capillary effect still dominates the hole development. Nevertheless, it is observed that the hole is formed closer to the rim. As a result, the expansion of the hole will interact with the rim and cause the rim to break \citep{Agbaglah_2021a}. 

\subsubsection{Impact of inlet gas turbulence on sheet breakup}
\begin{figure}
\centering
\includegraphics[trim={0cm 0cm 0cm 0cm},clip,width=1.\textwidth]{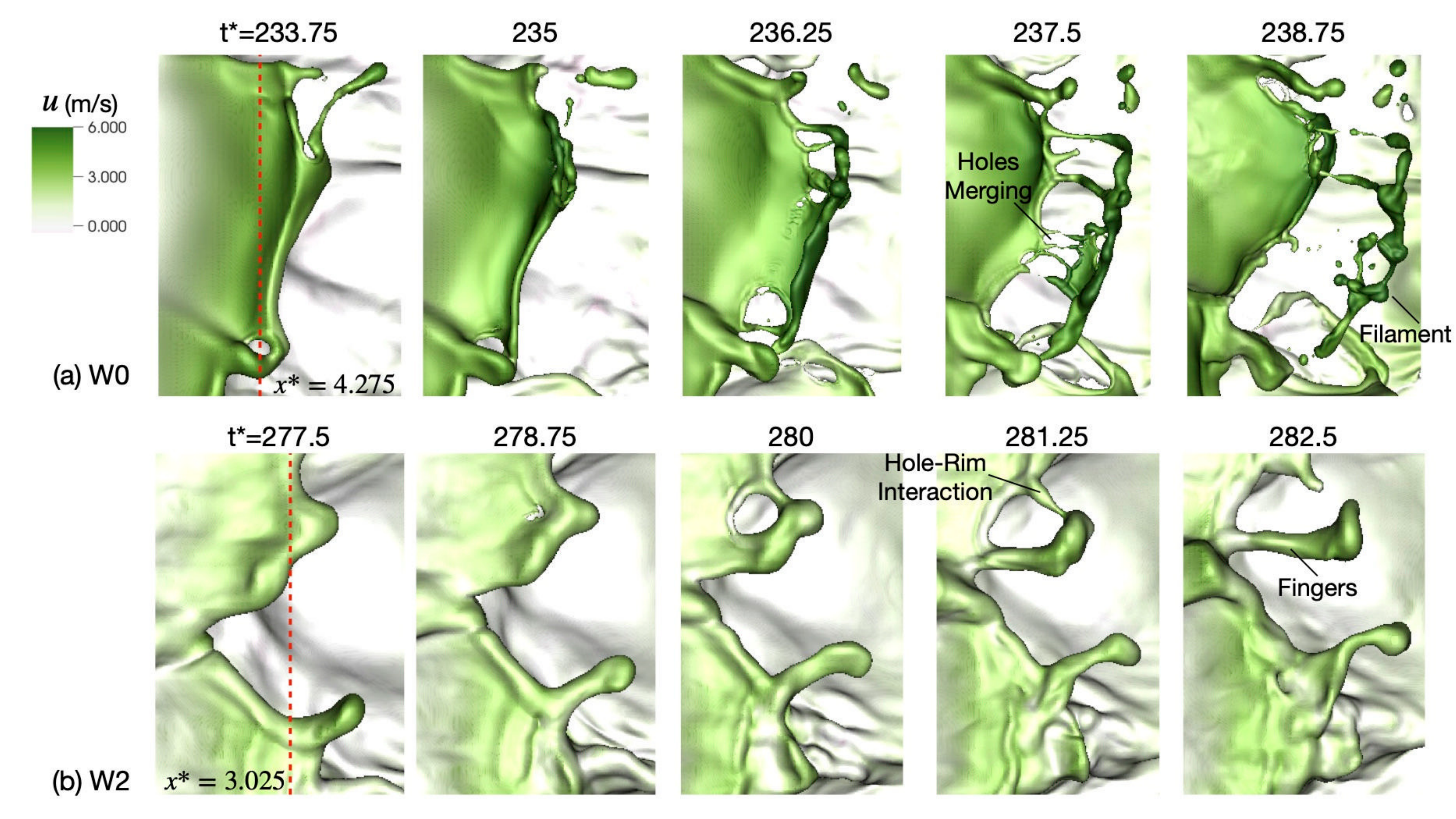}
\caption{Breakups of the liquid sheet due to hole formation and development for the cases W0 and W2. The interfaces are colored with $u$-velocity.}
\label{fig:sheet_breakup}
\end{figure}

%hole-rim interaction for turbulent, and hole-hole interaction for laminar \citep{Agbaglah_2021a}
Though the inlet gas turbulence does not change the hole dynamics, it does have a strong impact on the number of holes formed, and the locations where the holes are formed. The breakups of the liquid sheet for the W0 and W2 cases are shown in figure \ref{fig:sheet_breakup}. For the case W0, the holes are formed at around $x^*=4.275$, which is much further than the hole formation location $x^*=3.025$ for the turbulent gas inlet. This is due to smaller growth rate for the case W0, and as a result the interaction between the wave and the gas stream occur further downstream. Nevertheless, the interfacial wave for the case W0 reaches a larger amplitude and experiences a stronger interaction with the gas stream. Furthermore, the interfacial wave is accelerated by the gas stream to a higher velocity and is stretched to form a longer liquid sheet. The folding of the liquid sheet enhances the reduction of sheet thickness (see also figure \ref{fig:u_surf}). Multiple holes arise and expand simultaneously and the merging of the holes leads a violent breakup of the folding segment of the liquid sheet \citep{Agbaglah_2021a}. The rim of the liquid sheet is detached from the wave, forming a filament aligned with the transverse direction, see $t^*=238.75$ in figure \ref{fig:sheet_breakup}(a). The detached filament interacts with the gas stream and continues to break into droplets. Due to the irregular shape of the filament, the breakup dynamics is different from the regular Rayleigh breakup of a perturbed liquid cylinder \citep{Villermaux_2004a}. 

For the case W2, the transverse wavenumber is higher than that for the case W0. Therefore, the interfacial wave exhibits narrower lobes and more irregular rims. In the snapshots shown in figure \ref{fig:sheet_breakup}(b), only one hole is seen. By examining all the time snapshots, it is affirmed that the number of holes formed for the case W2 is significantly lower than that for the case W0. Since only one hole is formed, instead of merging with other holes as for the case W0, the hole expands and merges with the rim. The rim eventually breaks and retracts to form two fingers aligned with the longitudinal direction. The fingers are stretched by the gas stream and will break and form droplets. Since the size and orientation of the fingers/filaments formed for the case W2 are distinct from those for the case W0, the resulting droplet statistics will also be different. 

\subsection{Droplet statistics}
To obtain statistics of droplets formed in the interfacial wave breakup, the cells with liquid ($f>0$) that are connected together are tagged with the same ID. Then the volume and centroid coordinates of individual liquid structures can be computed. The details of the tagging approach can be found in previous study \citep{Ling_2015a}. 

\subsubsection{Size distribution}
\begin{figure}
\centering
\includegraphics[trim={0cm 0cm 0cm 0cm},clip,width=1\textwidth]{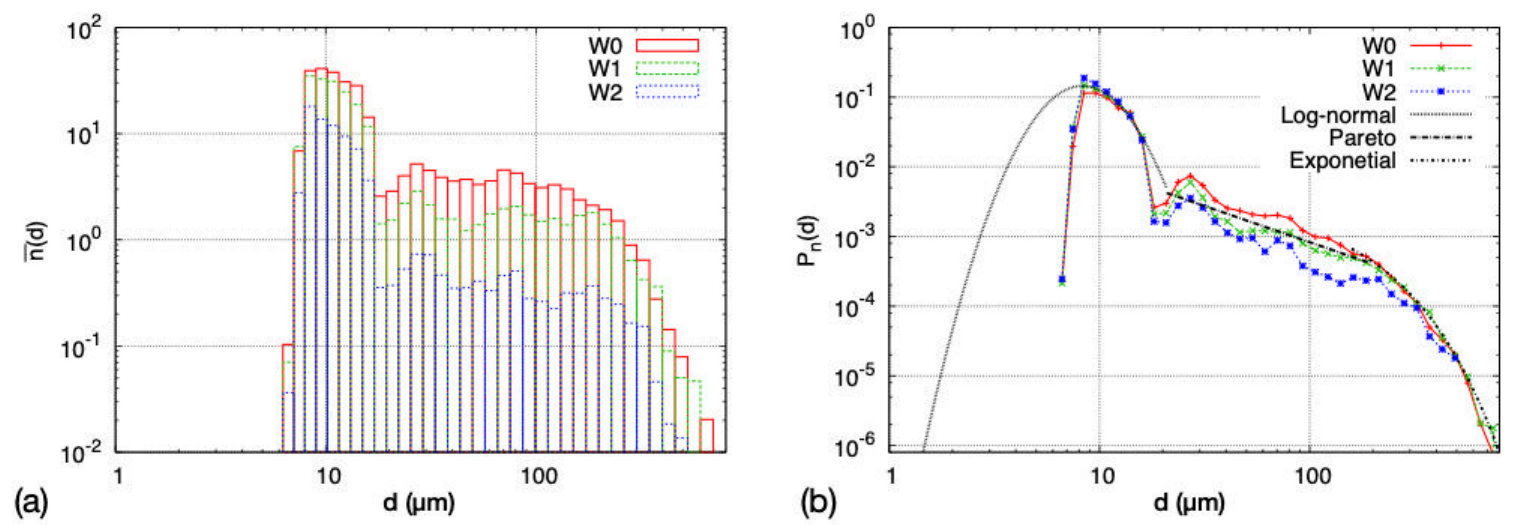}
\caption{(a) Droplet size distribution and (b) PDF based on the time-average droplet number for the cases W0, W1, and W2. }
\label{fig:droplet_stats}
\end{figure}

The droplet size distribution for different cases are shown in figure \ref{fig:droplet_stats}. Only the droplets in the region $x^*<14$ are considered, to exclude the effect of the outflow boundary. Furthermore, sampling is only performed after the droplet statistics has reached the statistically stationary state (at about $t^*=200$). For all cases, droplets are collected every 50 time steps, so in total about $N_s=$3000 samples have been collected. The sampling time has been verified to be sufficiently long to yield statistically converged results. Since the droplet number generally decreases with the droplet diameter $d$, the size interval $\Delta d$ is increased over $d$ with a scaling ratio 1.15, namely $\Delta d_{i+1} = 1.15 \Delta d_{i} $, where $i$ is the interval number. The number of droplets falling into the size interval centered with the diameter $d$ for all samples is denoted as $n$. The average droplet number is then defined as $\overline{n}=n/N_s$. 

The results for $\overline{n}$ as a function of $d$ for the cases W0, W1, and W2 are shown in figure \ref{fig:droplet_stats}(a). It is observed that $\overline{n}$ for all $d$ generally decreases when $I$ increases. This is consistent {with} the observation in figure \ref{fig:sheet_breakup}, namely the interfacial wave breakup is less violent when $I$ increases, and as a results, fewer droplets are formed. Nevertheless, the shapes of the distribution profiles for different $I$ are quite similar, which motivates the examination {of} the probability density function (PDF). The droplet size PDF based on the droplet number is defined as
\begin{equation}
	P_n(d)= \frac{\overline{n}(d)}{\Delta d\sum \overline{n}(d)}\, .
\end{equation}

{The droplet size PDF for different $I$ are shown in figure \ref{fig:droplet_stats}(b). The profile of $P_n$ is quite complex and cannot be represented by any typical distribution function. The recent study by \citet{Balachandar_2020a} indicates that a combined log-normal and Pareto distribution is required to represent the complex PDF for the respiratory droplets formed in a coughing event \citep{Duguid_1946a}. While the small droplets follow the log-normal distribution, the larger droplets follow the Pareto distribution. The drop size PDF for the present problem for different ranges of droplet size also seem to follow different distribution functions, \ie, the log-normal, Pareto, and exponential distribution functions. The expressions for these three functions are given as 
\begin{align}
	P_L(d) &= \frac{B_L}{{d \hat{\sigma} \sqrt{2\pi}}} \exp \left[ -\frac{(\ln d -\hat{\mu})^2}{2\hat{\sigma}^2}\right]\, ,
	\label{eq:log-normal}\\
	P_P(d) &= \frac{B_P}{d^\alpha}\, ,
	\label{eq:Pareto}\\
	P_E(d) &= B_E \exp(d/\lambda)\, ,
	\label{eq:Exponential}
\end{align}
where $B_L$, $B_P$, and $B_E$ are the corresponding normalization constants for the log-normal, Pareto, and exponential functions, respectively. For the log-normal function, $\hat{\mu}$ and $\hat{\sigma}$ are the expected value and the standard deviation of $\ln d$. While $\alpha$ is the power index for the Pareto distribution, $\lambda$ is the characteristic length for the exponential distribution. To compare the simulation results with the distribution functions, we have used the data for the case W1 to fit these distributions functions. For convenience of discussions, the droplets for $d \lesssim 25$ \textmu m, $25\ \lesssim d \lesssim 140$ \textmu m, and $d \gtrsim 140$ \textmu m are referred to as \emph{small}, \emph{medium} and \emph{large} droplets, the data of which are used to fit Eqs.\ \eqr{log-normal}, \eqr{Pareto}, and \eqr{Exponential}, respectively. The boundaries between different ranges vary slightly with $I$.  The values for the fitting parameters are $B_L=1.14,\ \hat{\sigma}=0.35,\ \hat{\mu}=2.24$ for the log-normal function, $B_P=0.099,\ \alpha=1.04$ for the Pareto function, and $B_E=0.0035,\ \lambda=95.0$ for the exponential function. The units for these parameters can be determined based on the units for $P_n$ and $d$, \ie, (\textmu m)$^{-1}$ and \textmu m, respectively. As shown in figure \ref{fig:droplet_stats}(b), the fitted distribution functions generally agree well with the simulation results.}

{
For the large droplets ($d \gtrsim 140$ \textmu m), the curves of $P_n$ for different $I$ approximately collapse to the same exponential function. The exponential tail in the drop size distribution has been observed in numerous experiments and simulations \citep{Simmons_1977a, Simmons_1977b, Marmottant_2004a, Ling_2015a}. 
For droplets of medium sizes, $25\ \lesssim d \lesssim 140$ \textmu m, the PDF exhibits a power-law decay, namely the Pareto distribution. The width of the Pareto distribution region increases with the  the inlet gas turbulence intensity $I$. As a result, the power-law decay is the most profound for the case W2. 
For the small droplets $d \lesssim 25$ \textmu m, the shape of $P_n$ is significantly different from that for the medium droplets. 
The different PDF profiles may be due to the different droplet formation mechanisms. While the large number of small droplets are probably generated directly in the breakup of liquid sheets, see figure \ref{fig:sheet_breakup}(a), the medium droplets are produced in the Rayleigh-Plateau breakup of thicker ligaments and fingers, see figure \ref{fig:sheet_breakup}(b). It will be interesting for future study to identify $P_n$ for different breakup events and to further reveal the physical reasons behind the different $P_n$ for the small and medium droplets. 
Following former studies \citep{Marty_2015a,Ling_2017a,Balachandar_2020a}, the log-normal function has been used to fit $P_n$ for the small droplets, though the narrow range of $d$ may have placed some uncertainty in the fitting process and whether the log-normal function is indeed the best fit for the small droplets. It should be noted that the droplets smaller than 8 \textmu m are not accounted here, since those droplets are smaller than about one cell and the formation of these droplets are likely under resolved. To better examine the distribution function for the small droplets, more data points are required for $d<8$ \textmu m, which would require further refined simulations. Such simulations require computer time at least an order of magnitude higher than that for the present simulations and will be relegated to future works. 
}

\subsubsection{Spatial distribution}
\begin{figure}
\centering
\includegraphics[trim={0cm 0cm 0cm 0cm},clip,width=1\textwidth]{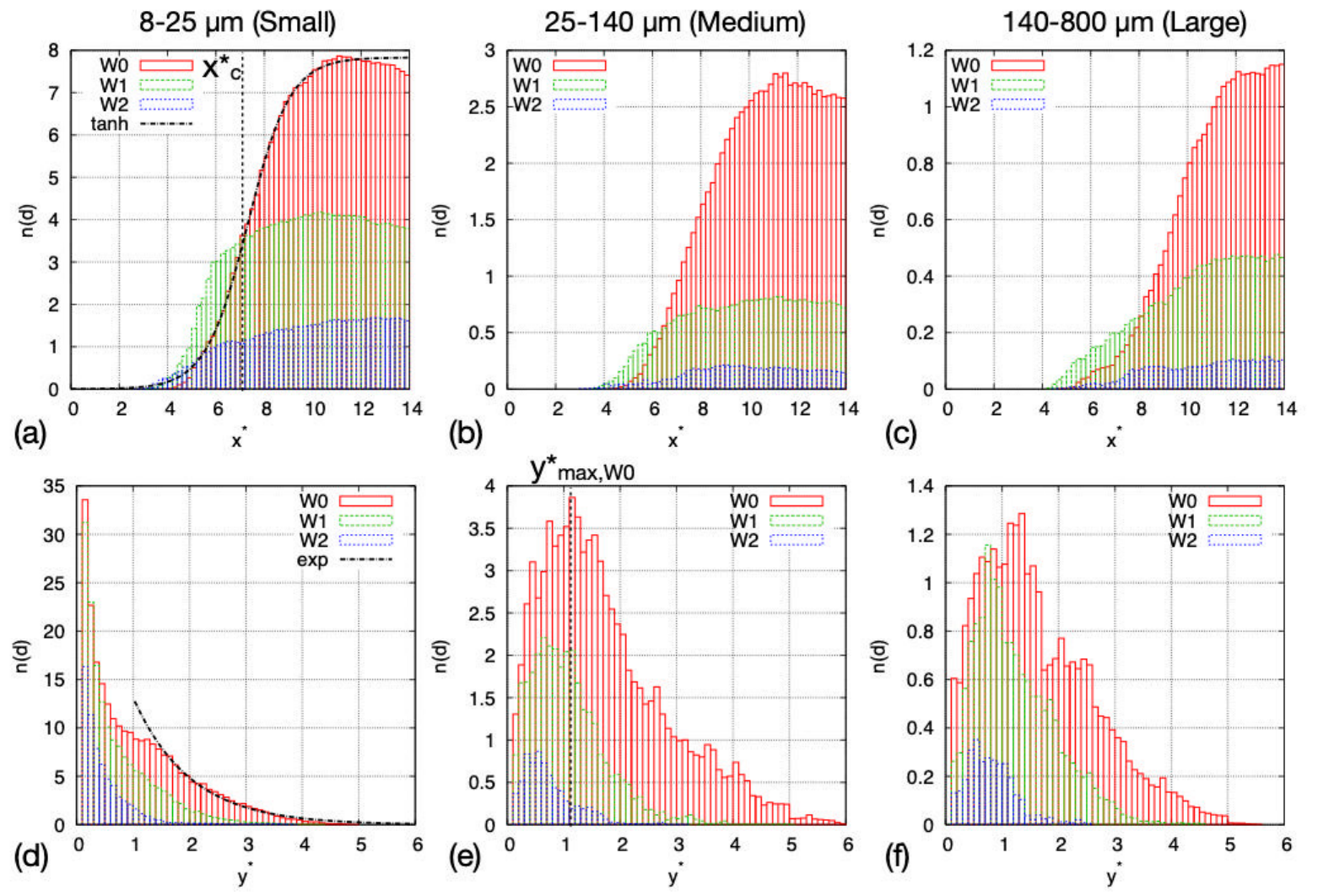}
\caption{Distribution of $\overline{n}$ in (a) streamwise and (b) vertical directions for the cases W0, W1, and W2. The droplets are collected every 50 time steps. The three columns are for the small (12.5--100 \textmu m), medium(100-- 200 \textmu m),  and large (200--300 \textmu m) droplets, respectively.}
\label{fig:droplet_sptial}
\end{figure}

The spatial distributions of droplet number $\overline{n}$ in the longitudinal and vertical directions for different droplet size ranges and the cases W0, W1, and W2 are shown in figure \ref{fig:droplet_sptial}. In the longitudinal direction, $\overline{n}$ is zero until the interfacial waves break. After that, $\overline{n}$ increases with $x^*$, approximately following the hyperbolic tangent function, see figure \ref{fig:droplet_sptial}(a). For the small and medium droplets, $\overline{n}$ reaches the maximum around $10\lesssim x^*\lesssim 13$. The decrease in $\overline{n}$ for $x^*\gtrsim 13$ is due to the coalescence of droplets and the merge of droplets back on the unbroken liquid layer at the bottom (see figure \ref{fig:general_behavior}{(c) and (d)}). For large droplets, the decrease of $\overline{n}$ is less profound, and $\overline{n}$ reaches a plateau for large $x^*$. When the inlet gas turbulence intensity $I$ increases, several changes are induced. First, the onset of the increase of $\overline{n}$ occurs at smaller $x^*$ due to the faster growth of the longitudinal wave and the earlier breakup of the interfacial waves. Second, the increase of $\overline{n}$ becomes more gradual and eventually settles on a lower plateau. As a result, the curves for W0 and W1 cross at a critical longitudinal location $x^*_c$. For $x^*<x^*_c$, $\overline{n}_\text{W1}> \overline{n}_\text{W0}$, or vice versa. The values of $x^*_c$ for the small and medium droplets are similar, at about $x^*_c=7$, while that for the large droplets is larger, at around $x^*_c=8.2$. 

The droplet number distributions in the vertical direction for different droplet size ranges are shown in figure \ref{fig:droplet_sptial}(d)-(f). For the small droplets, the maxima of $\overline{n}$ are at the bottom plane of domain for all cases. The droplet number then decreases rapidly with $y^*$ until $y^*\approx 1$, where the interface is located at the inlet and the decreases of $\overline{n}$ becomes more gradual. After that, $\overline{n}$ decreases approximately exponentially to zero, see figure \ref{fig:droplet_sptial}(d). The accumulation of small droplets near the bottom domain is probably due to the fact that the small droplets have smaller inertia and thus are easier to be carried by the gas stream toward the bottom  downstream. The medium and larger droplets have larger inertia and tend to maintain the velocity direction when they are formed. As a result, we see the maxima of $\overline{n}$ for the medium and the large droplets are near $y^*=1$, instead. The vertical location corresponding to the maximized $\overline{n}$ is denoted by $y^*_{\max}$ (see figure \ref{fig:droplet_sptial}(e)). It is observed that for the medium and large droplets, $y^*_{\max}$ decreases with the inlet gas turbulence intensity $I$. This is again due to the faster growth of the longitudinal wave for the cases with larger $I$, and as a result, the breakups of the interfacial waves occur in a lower vertical location (see also figure \ref{fig:general_behavior}).

\section{Conclusions}
\label{sec:conclusions}
The effect of inlet gas turbulence on the formation, development, and breakup of the interfacial waves in a two-phase mixing layer is investigated through direct numerical simulation. The gas and liquid properties and the mean injection velocities are chosen to place the present case in the absolute instability regime. Turbulent velocity fluctuations are induced at the gas inlet and a parametric study for the inlet gas turbulence intensity $I$ is performed. The governing equations are solved using the open-source multiphase flow solver, \emph{PARIS}, in which the mass-momentum consistent volume-of-fluid method is used to capture the sharp gas-liquid interfaces. To allow a detailed investigation of the transverse development of the interfacial waves, two different domain widths have been used. The wide domain consists of about 1.6 billions cells. 

The temporal evolutions and the frequency spectra for the interfacial height in the near field shows a dominant frequency, confirming that the case {is} in the absolute instability regime. The dominant frequency varies little when $I\lesssim 0.02$ but then increases over $I$ after the threshold. Similar increasing trend over $I$ has been observed for the spatial growth rate of the longitudinal wave amplitude. When the inlet gas is laminar, the dominant frequencies and spatial growth rates for the narrow and wide domains are similar. When the turbulence is present at the gas inlet, deviations between the results for the narrow and wide domains are observed. 

The longitudinal instability introduces vertical motion of the interface. When the interface moves upward and decelerates toward the gas, the Rayleigh-Taylor (RT) instability is triggered, introducing transverse modulations on the interfacial waves. The transverse variation is amplified as the wave interacts with the fast gas stream. The selection of dominant transverse wavenumber is dictated by the RT instability and can be determined by inviscid temporal stability theory with surface tension. The transverse wavenumber scales with the longitudinal frequency and thus will also increase with $I$. The simulation results for the transverse wavenumber agree well with theoretical predictions. The ratio between the growth rate of the transverse instability and the dominant frequency of the longitudinal instability is estimated to be comparable, affirming the RT instability is responsible for the initiation of the transverse modulations for the present problem.

The subsequent development of the interfacial wave is controlled by the nonlinear interaction with the gas stream. When the interfacial wave invades in the gas stream, an aerodynamic drag is developed, which further destabilizes the interfacial waves and stretches the wave crest to form liquid lobes/sheets. For the laminar gas inlet, the liquid sheet deforms and folds when it interacts with the gas stream, leading {to the} formation of multiple holes in the folding area. The expansion and merging of the holes eventually lead to a violent disintegration of the liquid sheet, detaching filaments that are aligned with the transverse direction. In contrast, for the turbulent gas inlet, the lobes are narrower and the rim is less regular. Fewer holes are formed and the expansion of the holes interacts with the rim, forming fingers aligned with the longitudinal direction. The different breakup dynamics of the liquid sheet results in different sizes and orientations of the filaments. The droplets formed due to the subsequent breakup of the filaments thus exhibit different statistics. 

The numbers of droplets for all sizes reduce when $I$ increases, though the probability density functions (PDF) appear to be similar. The droplets of different size ranges are found to follow different distribution functions, \ie, the log-normal, Pareto, and exponential functions for the small ($8\ \lesssim d \lesssim 25$ \textmu m), medium ($25\ \lesssim d \lesssim 140$ \textmu m), and large ($d \gtrsim 140$ \textmu m) droplets, respectively. The spatial distributions of the droplet number for different size ranges have also been presented. The increase of the droplet number along the longitudinal direction is similar to the hyperbolic tangent function and the decrease of droplet number along the vertical distance away from the mixing layer follows the exponential function. When $I$ increases, more droplets are formed in the near field and near the bottom of the domain.

\section*{Acknowledgement}
This research is supported by the National Science Foundation (\#1942324). The authors also acknowledge the Extreme Science and Engineering Discovery Environment (XSEDE) for providing the computational resources that have contributed to the research results reported in this paper. The Baylor High Performance and Research Computing Services (HPRCS) have been used to process the simulation data. We also thank Dr. Gretar Tryggvason, Dr.~Stephane Zaleski, and other developers of for their contributions to the \emph{PARIS} solver.

\section*{Declaration of Interests}
The authors report no conflict of interest.

%\bibliographystyle{jfm}
%\bibliography{refs}

\begin{thebibliography}{62}
\expandafter\ifx\csname natexlab\endcsname\relax\def\natexlab#1{#1}\fi
\def\au#1{#1} \def\ed#1{#1} \def\yr#1{#1}\def\at#1{#1}\def\jt#1{\textit{#1}}
  \def\bt#1{#1}\def\bvol#1{\textbf{#1}} \def\vol#1{#1} \def\pg#1{#1}
  \def\publ#1{#1}\def\arxiv#1{#1}\def\org#1{#1}\def\st#1{\textit{#1}}

\bibitem[Agbaglah {\em et~al.\/}(2017)Agbaglah, Chiodi \&
  Desjardins]{Agbaglah_2017a}
{\sc \au{Agbaglah, G.}, \au{Chiodi, R.} \& \au{Desjardins, O.}} \yr{2017}
  \at{Numerical simulation of the initial destabilization of an air-blasted
  liquid layer}.  \jt{J. Fluid Mech.}  \bvol{812},  \pg{1024--1038}.

\bibitem[Agbaglah {\em et~al.\/}(2013)Agbaglah, Josserand \&
  Zaleski]{Agbaglah_2013a}
{\sc \au{Agbaglah, G.}, \au{Josserand, C.} \& \au{Zaleski, S.}} \yr{2013}
  \at{Longitudinal instability of a liquid rim}.  \jt{Phys.~Fluids}  \bvol{25},
   \pg{022103}.

\bibitem[Agbaglah(2021)]{Agbaglah_2021a}
{\sc \au{Agbaglah, G.~G.}} \yr{2021}  \at{Breakup of thin liquid sheets through
  hole--hole and hole--rim merging}.  \jt{J.~Fluid Mech.}  \bvol{911},
  \pg{A23}.

\bibitem[Ambravaneswaran {\em et~al.\/}(2002)Ambravaneswaran, Wilkes \&
  Basaran]{Ambravaneswaran_2002a}
{\sc \au{Ambravaneswaran, B.}, \au{Wilkes, E.~D.} \& \au{Basaran, O.~A.}}
  \yr{2002}  \at{Drop formation from a capillary tube: Comparison of
  one-dimensional and two-dimensional analyses and occurrence of satellite
  drops}.  \jt{Phys.~Fluids}  \bvol{14},  \pg{2606--2621}.

\bibitem[Aniszewski {\em et~al.\/}(2021)Aniszewski, Arrufat, Crialesi-Esposito,
  Dabiri, Fuster, Ling, Lu, Malan, Pal, Scardovelli, Tryggvason, Yecko \&
  Zaleski]{Aniszewski_2021a}
{\sc \au{Aniszewski, W.}, \au{Arrufat, T.}, \au{Crialesi-Esposito, M.},
  \au{Dabiri, S.}, \au{Fuster, D.}, \au{Ling, Y.}, \au{Lu, J.}, \au{Malan, L.},
  \au{Pal, S.}, \au{Scardovelli, R.}, \au{Tryggvason, G.}, \au{Yecko, P.} \&
  \au{Zaleski, S.}} \yr{2021}  \at{{PArallel, Robust, Interface Simulator
  (PARIS)}}.  \jt{Comput.~Phys.~Comm.}  \bvol{263},  \pg{107849}.

\bibitem[Arrufat {\em et~al.\/}(2020)Arrufat, Crialesi-Esposito, Fuster, Ling,
  Malan, Pal, Scardovelli, Tryggvason \& Zaleski]{Arrufat_2020a}
{\sc \au{Arrufat, T.}, \au{Crialesi-Esposito, M.}, \au{Fuster, D.}, \au{Ling,
  Y.}, \au{Malan, L.}, \au{Pal, S.}, \au{Scardovelli, R.}, \au{Tryggvason, G.}
  \& \au{Zaleski, S.}} \yr{2020}  \at{A momentum-conserving, consistent,
  volume-of-fluid method for incompressible flow on staggered grids}.
  \jt{Comput.~Fluids}  \bvol{215},  \pg{104785}.

\bibitem[Aulisa {\em et~al.\/}(2007)Aulisa, Manservisi, Scardovelli \&
  Zaleski]{Aulisa_2007a}
{\sc \au{Aulisa, E.}, \au{Manservisi, S.}, \au{Scardovelli, R.} \& \au{Zaleski,
  S.}} \yr{2007}  \at{Interface reconstruction with least-squares fit and split
  advection in three-dimensional cartesian geometry}.  \jt{J.~Comput.~Phys.}
  \bvol{225},  \pg{2301--2319}.

\bibitem[Balachandar {\em et~al.\/}(2020)Balachandar, Zaleski, Soldati, Ahmadi
  \& Bourouiba]{Balachandar_2020a}
{\sc \au{Balachandar, S.}, \au{Zaleski, S.}, \au{Soldati, A.}, \au{Ahmadi, G.}
  \& \au{Bourouiba, L.}} \yr{2020}  \at{Host-to-host airborne transmission as a
  multiphase flow problem for science-based social distance guidelines}.
  \jt{Int.~J.~Multiphase Flow}  \bvol{132},  \pg{103439}.

\bibitem[Boeck \& Zaleski(2005)]{Boeck_2005a}
{\sc \au{Boeck, T.} \& \au{Zaleski, S.}} \yr{2005}  \at{Viscous versus inviscid
  instability of two-phase mixing layers with continuous velocity profile}.
  \jt{Phys.~Fluids}  \bvol{17},  \pg{032106}.

\bibitem[Castrej{\'o}n-Pita {\em et~al.\/}(2015)Castrej{\'o}n-Pita,
  Castrej{\'o}n-Pita, Thete, Sambath, Hutchings, Hinch, Lister \&
  Basaran]{Castrejon-Pita_2015a}
{\sc \au{Castrej{\'o}n-Pita, J.~R.}, \au{Castrej{\'o}n-Pita, A.~A.}, \au{Thete,
  S.~S.}, \au{Sambath, K.}, \au{Hutchings, I.~M.}, \au{Hinch, J.}, \au{Lister,
  J.~R.} \& \au{Basaran, O.~A.}} \yr{2015}  \at{Plethora of transitions during
  breakup of liquid filaments}.  \jt{Proc. Natl. Acad. Sci. U.S.A.}
  \bvol{112},  \pg{4582--4587}.

\bibitem[Chandrasekhar(1961)]{Chandrasekhar_1961a}
{\sc \au{Chandrasekhar, S.}} \yr{1961} {\em Hydrodynamic and hydromagnetic
  stability\/}.  \publ{Oxford University Press.}

\bibitem[Chaussonnet {\em et~al.\/}(2020)Chaussonnet, Gepperth, Holz, Koch \&
  Bauer]{Chaussonnet_2020a}
{\sc \au{Chaussonnet, G.}, \au{Gepperth, S.}, \au{Holz, S.}, \au{Koch, R.} \&
  \au{Bauer, H.-J.}} \yr{2020}  \at{Influence of the ambient pressure on the
  liquid accumulation and on the primary spray in prefilming airblast
  atomization}.  \jt{Int.~J.~Multiphase Flow}  \bvol{125},  \pg{103229}.

\bibitem[Chorin(1968)]{Chorin_1968a}
{\sc \au{Chorin, A.~J.}} \yr{1968}  \at{Numerical solution of the
  {Navier-Stokes} equations}.  \jt{Math.~Comput.}  \bvol{22},  \pg{745--762}.

\bibitem[Delon {\em et~al.\/}(2018)Delon, Cartellier \& Matas]{Delon_2018a}
{\sc \au{Delon, A.}, \au{Cartellier, A.} \& \au{Matas, J.-P.}} \yr{2018}
  \at{Flapping instability of a liquid jet}.  \jt{Phys.~Rev.~Fluids}  \bvol{3},
   \pg{043901}.

\bibitem[Dimotakis(1986)]{Dimotakis_1986a}
{\sc \au{Dimotakis, P.~E.}} \yr{1986}  \at{Two-dimensional shear-layer
  entrainment}.  \jt{{AIAA} J.}  \bvol{24},  \pg{1791--1796}.

\bibitem[Duguid(1946)]{Duguid_1946a}
{\sc \au{Duguid, J.~P.}} \yr{1946}  \at{The size and the duration of
  air-carriage of respiratory droplets and droplet-nuclei}.
  \jt{Epidemiol.~Infect.}  \bvol{44},  \pg{471--479}.

\bibitem[Eggers(1993)]{Eggers_1993a}
{\sc \au{Eggers, J.}} \yr{1993}  \at{Universal pinching of {3D} axisymmetric
  free-surface flow}.  \jt{Phys.~Rev.~Lett.}  \bvol{71},  \pg{3458}.

\bibitem[Francois {\em et~al.\/}(2006)Francois, Cummins, Dendy, Kothe, Sicilian
  \& Williams]{Francois_2006a}
{\sc \au{Francois, M.~M.}, \au{Cummins, S.~J.}, \au{Dendy, E.~D.}, \au{Kothe,
  D.~B.}, \au{Sicilian, J.~M.} \& \au{Williams, M.~W.}} \yr{2006}  \at{A
  balanced-force algorithm for continuous and sharp interfacial surface tension
  models within a volume tracking framework}.  \jt{J.~Comput.~Phys.}
  \bvol{213},  \pg{141--173}.

\bibitem[Fuster {\em et~al.\/}(2013)Fuster, Matas, Marty, Popinet, J.,
  Cartellier \& Zaleski]{Fuster_2013a}
{\sc \au{Fuster, D.}, \au{Matas, J.~P.}, \au{Marty, S.}, \au{Popinet, S.},
  \au{J., Hoepffner}, \au{Cartellier, A.} \& \au{Zaleski, S.}} \yr{2013}
  \at{Instability regimes in the primary breakup region of planar coflowing
  sheets}.  \jt{J. Fluid Mech}  \bvol{736},  \pg{150--176}.

\bibitem[Healey(2007)]{Healey_2007a}
{\sc \au{Healey, J.~J.}} \yr{2007}  \at{Enhancing the absolute instability of a
  boundary layer by adding a far-away plate}.  \jt{J.~Fluid Mech.}  \bvol{579},
   \pg{29}.

\bibitem[Herrmann(2011)]{Herrmann_2011a}
{\sc \au{Herrmann, M.}} \yr{2011}  \at{The influence of density ratio on the
  primary atomization of a turbulent liquid jet in crossflow}.
  \jt{Proc.~Combust.~Inst.}  \bvol{33},  \pg{2079--2088}.

\bibitem[Hoepffner {\em et~al.\/}(2011)Hoepffner, Blumenthal \&
  Zaleski]{Hoepffner_2011a}
{\sc \au{Hoepffner, J.}, \au{Blumenthal, R.} \& \au{Zaleski, S.}} \yr{2011}
  \at{Self-similar wave produced by local perturbation of the
  {Kelvin-Helmholtz} shear-layer instability}.  \jt{Phys.~Rev.~Lett.}
  \bvol{106},  \pg{104502}.

\bibitem[Jarrahbashi \& Sirignano(2014)]{Jarrahbashi_2014a}
{\sc \au{Jarrahbashi, D.} \& \au{Sirignano, W.~A.}} \yr{2014}  \at{Vorticity
  dynamics for transient high-pressure liquid injection a}.  \jt{Phys.~Fluids}
  \bvol{26},  \pg{73}.

\bibitem[Jarrahbashi {\em et~al.\/}(2016)Jarrahbashi, Sirignano, Popov \&
  Hussain]{Jarrahbashi_2016a}
{\sc \au{Jarrahbashi, D.}, \au{Sirignano, W.~A.}, \au{Popov, P.~P.} \&
  \au{Hussain, F.}} \yr{2016}  \at{Early spray development at high gas density:
  hole, ligament and bridge formations}.  \jt{J. Fluid Mech.}  \bvol{792},
  \pg{186--231}.

\bibitem[Jiang \& Ling(2020)]{Jiang_2020a}
{\sc \au{Jiang, D.} \& \au{Ling, Y.}} \yr{2020}  \at{Destabilization of a
  planar liquid stream by a co-flowing turbulent gas stream}.
  \jt{Int.~J.~Multiphase Flow}  \bvol{122},  \pg{103121}.

\bibitem[Joseph {\em et~al.\/}(1999)Joseph, Belanger \& Beavers]{Joseph_1999a}
{\sc \au{Joseph, D.~D.}, \au{Belanger, J.} \& \au{Beavers, G.~S.}} \yr{1999}
  \at{Breakup of a liquid drop suddenly exposed to a high-speed airstream}.
  \jt{Int.~J.~Multiphase Flow}  \bvol{25},  \pg{1263--1303}.

\bibitem[Juniper(2006)]{Juniper_2006a}
{\sc \au{Juniper, M.~P}} \yr{2006}  \at{The effect of confinement on the
  stability of two-dimensional shear flows}.  \jt{J. Fluid Mech.}  \bvol{565},
  \pg{171--195}.

\bibitem[Klein {\em et~al.\/}(2003)Klein, Sadiki \& Janicka]{Klein_2003a}
{\sc \au{Klein, M.}, \au{Sadiki, A.} \& \au{Janicka, J.}} \yr{2003}  \at{A
  digital filter based generation of inflow data for spatially developing
  direct numerical or large eddy simulations}.  \jt{J.~Comput.~Phys.}
  \bvol{186},  \pg{652--665}.

\bibitem[Kooij {\em et~al.\/}(2018)Kooij, Sijs, Denn, Villermaux \&
  Bonn]{Kooij_2018a}
{\sc \au{Kooij, S.}, \au{Sijs, R.}, \au{Denn, M.~M.}, \au{Villermaux, E.} \&
  \au{Bonn, D.}} \yr{2018}  \at{What determines the drop size in sprays?}
  \jt{Phys.~Rev.~X}  \bvol{8},  \pg{031019}.

\bibitem[Lasheras {\em et~al.\/}(1998)Lasheras, Villermaux \&
  Hopfinger]{Lasheras_1998a}
{\sc \au{Lasheras, J.~C.}, \au{Villermaux, E.} \& \au{Hopfinger, E.~J.}}
  \yr{1998}  \at{Break-up and atomization of a round water jet by a high-speed
  annular air jet}.  \jt{J.~Fluid Mech.}  \bvol{357},  \pg{351--379}.

\bibitem[Lefebvre(1988)]{Lefebvre_1988a}
{\sc \au{Lefebvre, A.}} \yr{1988} {\em Atomization and sprays\/}.  \publ{CRC
  press}.

\bibitem[Lefebvre(1980)]{Lefebvre_1980a}
{\sc \au{Lefebvre, A.~H.}} \yr{1980}  \at{Airblast atomization}.
  \jt{Prog.~Energ.~Combust.~Sci.}  \bvol{6},  \pg{233--261}.

\bibitem[Li(1995)]{Li_1995b}
{\sc \au{Li, J.}} \yr{1995}  \at{Calcul d'interface affine par morceaux
  (piecewise linear interface calculation)}.  \jt{C. R. Acad. Sci. Paris,
  s{\'e}rie {II b}}  \bvol{320},  \pg{391--396}.

\bibitem[Ling {\em et~al.\/}(2019)Ling, Fuster, Tryggvasson \&
  Zaleski]{Ling_2019a}
{\sc \au{Ling, Y.}, \au{Fuster, D.}, \au{Tryggvasson, G.} \& \au{Zaleski, S.}}
  \yr{2019}  \at{A two-phase mixing layer between parallel gas and liquid
  streams: multiphase turbulence statistics and influence of interfacial
  instability}.  \jt{J.~Fluid Mech.}  \bvol{859},  \pg{268--307}.

\bibitem[Ling {\em et~al.\/}(2017)Ling, Fuster, Zaleski \&
  Tryggvason]{Ling_2017a}
{\sc \au{Ling, Y.}, \au{Fuster, D.}, \au{Zaleski, S.} \& \au{Tryggvason, G.}}
  \yr{2017}  \at{Spray formation in a quasiplanar gas-liquid mixing layer at
  moderate density ratios: A numerical closeup}.  \jt{Phys.~Rev.~Fluids}
  \bvol{2},  \pg{014005}.

\bibitem[Ling {\em et~al.\/}(2015)Ling, Zaleski \& Scardovelli]{Ling_2015a}
{\sc \au{Ling, Y.}, \au{Zaleski, S.} \& \au{Scardovelli, R.}} \yr{2015}
  \at{Multiscale simulation of atomization with small droplets represented by a
  lagrangian point-particle model}.  \jt{Int.~J.~Multiphase Flow}  \bvol{76},
  \pg{122--143}.

\bibitem[Marmottant \& Villermaux(2004)]{Marmottant_2004a}
{\sc \au{Marmottant, P.} \& \au{Villermaux, E.}} \yr{2004}  \at{On spray
  formation}.  \jt{J.~Fluid Mech.}  \bvol{498},  \pg{73--111}.

\bibitem[Marston {\em et~al.\/}(2016)Marston, Truscott, Speirs, Mansoor \&
  Thoroddsen]{Marston_2016a}
{\sc \au{Marston, J.~O.}, \au{Truscott, T.~T.}, \au{Speirs, N.~B.},
  \au{Mansoor, M.~M.} \& \au{Thoroddsen, S.~T.}} \yr{2016}  \at{Crown sealing
  and buckling instability during water entry of spheres}.  \jt{J. Fluid Mech.}
   \bvol{794},  \pg{506--529}.

\bibitem[Marty(2015)]{Marty_2015a}
{\sc \au{Marty, S.}} \yr{2015}  \at{Contribution a l'\'{e}tude de l'atomisation
  assist\'{e}e d'un liquide}. PhD thesis, Universit\'{e} de Grenoble.

\bibitem[Matas(2015)]{Matas_2015b}
{\sc \au{Matas, J.-P.}} \yr{2015}  \at{Inviscid versus viscous instability
  mechanism of an air--water mixing layer}.  \jt{J. Fluid Mech.}  \bvol{768},
  \pg{375--387}.

\bibitem[Matas {\em et~al.\/}(2018)Matas, Delon \& Cartellier]{Matas_2018a}
{\sc \au{Matas, J.-P.}, \au{Delon, A.} \& \au{Cartellier, A.}} \yr{2018}
  \at{Shear instability of an axisymmetric air--water coaxial jet}.
  \jt{J.~Fluid Mech.}  \bvol{843},  \pg{575--600}.

\bibitem[Matas {\em et~al.\/}(2011)Matas, Marty \& Cartellier]{Matas_2011a}
{\sc \au{Matas, J.-P.}, \au{Marty, S.} \& \au{Cartellier, A.}} \yr{2011}
  \at{Experimental and analytical study of the shear instability of a
  gas-liquid mixing layer}.  \jt{Phys.~Fluids}  \bvol{23},  \pg{094112}.

\bibitem[Matas {\em et~al.\/}(2015)Matas, Marty, Dem \&
  Cartellier]{Matas_2015a}
{\sc \au{Matas, J.-P.}, \au{Marty, S.}, \au{Dem, M.~S.} \& \au{Cartellier, A.}}
  \yr{2015}  \at{Influence of gas turbulence on the instability of an air-water
  mixing layer}.  \jt{Phys.~Rev.~Lett.}  \bvol{115},  \pg{074501}.

\bibitem[O'Naraigh {\em et~al.\/}(2013)O'Naraigh, Spelt \& Shaw]{Naraigh_2013a}
{\sc \au{O'Naraigh, L.}, \au{Spelt, P. D.~M.} \& \au{Shaw, S.~J.}} \yr{2013}
  \at{Absolute linear instability in laminar and turbulent gas--liquid
  two-layer channel flow}.  \jt{J. Fluid Mech.}  \bvol{714},  \pg{58--94}.

\bibitem[Opfer {\em et~al.\/}(2014)Opfer, Roisman, Venzmer, Klostermann \&
  Tropea]{Opfer_2014a}
{\sc \au{Opfer, L.}, \au{Roisman, I.~V.}, \au{Venzmer, J.}, \au{Klostermann,
  M.} \& \au{Tropea, C.}} \yr{2014}  \at{Droplet-air collision dynamics:
  Evolution of the film thickness}.  \jt{Phys.~Rev.~E}  \bvol{89},
  \pg{013023}.

\bibitem[Otto {\em et~al.\/}(2013)Otto, Rossi \& Boeck]{Otto_2013a}
{\sc \au{Otto, T.}, \au{Rossi, M.} \& \au{Boeck, T.}} \yr{2013}  \at{Viscous
  instability of a sheared liquid-gas interface: Dependence on fluid properties
  and basic velocity profile}.  \jt{Phys.~Fluids}  \bvol{25},  \pg{032103}.

\bibitem[Popinet(2009)]{Popinet_2009a}
{\sc \au{Popinet, S.}} \yr{2009}  \at{An accurate adaptive solver for
  surface-tension-driven interfacial flows}.  \jt{J.~Comput.~Phys.}
  \bvol{228}~(16),  \pg{5838--5866}.

\bibitem[Raynal(1997)]{Raynal_1997a}
{\sc \au{Raynal, L.}} \yr{1997}  \at{Instabilite et entrainement a l'interface
  d'une couche de melange liquide-gaz}. PhD thesis, Universit\'{e} Joseph
  Fourier - Grenoble I.

\bibitem[Renardy \& Renardy(2002)]{Renardy_2002a}
{\sc \au{Renardy, Y.} \& \au{Renardy, M.}} \yr{2002}  \at{{PROST}: a parabolic
  reconstruction of surface tension for the volume-of-fluid method}.
  \jt{J.~Comput.~Phys.}  \bvol{183},  \pg{400--421}.

\bibitem[Roisman(2010)]{Roisman_2010a}
{\sc \au{Roisman, I.~V.}} \yr{2010}  \at{On the instability of a free viscous
  rim}.  \jt{J. Fluid Mech.}  \bvol{661},  \pg{206--228}.

\bibitem[Rudman(1998)]{Rudman_1998a}
{\sc \au{Rudman, M.}} \yr{1998}  \at{A volume-tracking method for
  incompressible multifluid flows with large density variations}.
  \jt{Int.~J.~Numer.~Meth.~Fluids}  \bvol{28},  \pg{357--378}.

\bibitem[Scardovelli \& Zaleski(2003)]{Scardovelli_2003a}
{\sc \au{Scardovelli, R.} \& \au{Zaleski, S.}} \yr{2003}  \at{Interface
  reconstruction with least-square fit and split eulerian--lagrangian
  advection}.  \jt{Int.~J.~Numer.~Meth.~Fluids}  \bvol{41}~(3),  \pg{251--274}.

\bibitem[Simmons(1977{\natexlab{{\em a\/}}})]{Simmons_1977a}
{\sc \au{Simmons, H.~C.}} \yr{1977{\natexlab{{\em a\/}}}}  \at{The correlation
  of drop-size distributions in fuel nozzle sprays---part i: The
  drop-size/volume-fraction distribution}.  \jt{J. Engng.~Power}  \bvol{7},
  \pg{309--314}.

\bibitem[Simmons(1977{\natexlab{{\em b\/}}})]{Simmons_1977b}
{\sc \au{Simmons, H.~C.}} \yr{1977{\natexlab{{\em b\/}}}}  \at{The correlation
  of drop-size distributions in fuel nozzle sprays---part ii: The
  drop-size/number distribution}.  \jt{J. Engng.~Power}  \bvol{7},
  \pg{315--319}.

\bibitem[Snoeijer \& Andreotti(2013)]{Snoeijer_2013a}
{\sc \au{Snoeijer, J.~H.} \& \au{Andreotti, B.}} \yr{2013}  \at{Moving contact
  lines: scales, regimes, and dynamical transitions}.  \jt{Annu.~Rev.~Fluid
  Mech.}  \bvol{45},  \pg{269--292}.

\bibitem[Tryggvason {\em et~al.\/}(2011)Tryggvason, Scardovelli \&
  Zaleski]{Tryggvason_2011a}
{\sc \au{Tryggvason, G.}, \au{Scardovelli, R.} \& \au{Zaleski, S.}} \yr{2011}
  {\em Direct numerical simulations of gas-liquid multiphase flows\/}.
  \publ{Cambridge University Press}.

\bibitem[Varga {\em et~al.\/}(2003)Varga, Lasheras \& Hopfinger]{Varga_2003a}
{\sc \au{Varga, C.~M.}, \au{Lasheras, J.~C.} \& \au{Hopfinger, E.~J.}}
  \yr{2003}  \at{Initial breakup of a small-diameter liquid jet by a high-speed
  gas stream}.  \jt{J.~Fluid Mech.}  \bvol{497},  \pg{405--434}.

\bibitem[Vaudor {\em et~al.\/}(2017)Vaudor, M\'enard, Aniszewski, Doring \&
  Berlemont]{Vaudor_2017a}
{\sc \au{Vaudor, G.}, \au{M\'enard, T.}, \au{Aniszewski, W.}, \au{Doring, M.}
  \& \au{Berlemont, A.}} \yr{2017}  \at{A consistent mass and momentum flux
  computation method for two phase flows. {Application} to atomization
  process}.  \jt{Comput.~Fluids}  \bvol{152},  \pg{204--216}.

\bibitem[Villermaux {\em et~al.\/}(2004)Villermaux, Marmottant \&
  Duplat]{Villermaux_2004a}
{\sc \au{Villermaux, E.}, \au{Marmottant, Ph.} \& \au{Duplat, J.}} \yr{2004}
  \at{Ligament-mediated spray formation}.  \jt{Phys.~Rev.~Lett.}  \bvol{92},
  \pg{074501}.

\bibitem[Zandian {\em et~al.\/}(2018)Zandian, Sirignano \&
  Hussain]{Zandian_2018a}
{\sc \au{Zandian, A.}, \au{Sirignano, W.~A.} \& \au{Hussain, F.}} \yr{2018}
  \at{Understanding liquid-jet atomization cascades via vortex dynamics}.
  \jt{J.~Fluid Mech.}  \bvol{843},  \pg{293--354}.

\bibitem[Zhang {\em et~al.\/}(2019)Zhang, Ling, Tsai, Wang, Popinet \&
  Zaleski]{Zhang_2019b}
{\sc \au{Zhang, B.}, \au{Ling, Y.}, \au{Tsai, P.-H.}, \au{Wang, A.-B.},
  \au{Popinet, S.} \& \au{Zaleski, S.}} \yr{2019}  \at{Short-term oscillation
  and falling dynamics for a water drop dripping in quiescent air}.
  \jt{Phys.~Rev.~Fluids}  \bvol{4},  \pg{123604}.

\bibitem[Zhang {\em et~al.\/}(2020)Zhang, Popinet \& Ling]{Zhang_2020a}
{\sc \au{Zhang, B.}, \au{Popinet, S.} \& \au{Ling, Y.}} \yr{2020}  \at{Modeling
  and detailed numerical simulation of the primary breakup of a gasoline
  surrogate jet under non-evaporative operating conditions}.
  \jt{Int.~J.~Multiphase Flow}  \bvol{130},  \pg{103362}.

\end{thebibliography}

\end{document}